\documentclass[aps,twocolumn,floatfix,showpacs,amssymb]{revtex4}
\usepackage{graphicx,bm}
\usepackage{hyperref}
\usepackage{amsmath}
\usepackage{mathbbol}
\usepackage{color}

\begin{document}
\title{Second-order virial expansion for an atomic gas in a harmonic waveguide}
\author{Tom Kristensen$^{1,2,3}$, Xavier Leyronas$^{4}$ and Ludovic Pricoupenko$^{1,2}$}

\affiliation
{
1- Sorbonne Universit\'{e}s, UPMC Univ Paris 06, UMR 7600, Laboratoire de Physique Th\'{e}orique de la Mati\`{e}re Condens\'{e}e, F-75005, Paris, France\\
2- CNRS, UMR 7600, Laboratoire de Physique Th\'{e}orique de la Mati\`{e}re Condens\'{e}e, F-75005, Paris, France\\
3-Institut de Physique de Rennes, UMR 6251 du CNRS and Universit\'{e} de Rennes 1, 35042 Rennes Cedex, France\\
4- Laboratoire de Physique Statistique, \'{E}cole Normale
Sup\'{e}rieure, PSL Research University; Universit\'{e} Paris Diderot Sorbonne Paris-Cit\'{e}; Sorbonne Universit\'{e}s UPMC Univ Paris 06; CNRS; 24 rue Lhomond, 75005 Paris, France.
}
\date{\today}


\begin{abstract}
The virial expansion for cold two-component Fermi and Bose atomic gases is considered in the presence of a waveguide and in the vicinity of a Feshbach resonance. The interaction between atoms and the coupling with the Feshbach molecules is modeled using a quantitative separable two-channel model. The scattering phase-shift in an atomic waveguide is defined. This permits us to extend the Beth-Uhlenbeck formula for the second-order virial coefficient to this inhomogeneous case.
\end{abstract}

\maketitle

\section*{INTRODUCTION}

The use of magnetic Feshbach resonances permits one to achieve the unitary regime in atomic gases where universal thermodynamic properties such as the equation of state (EOS) are expected \cite{Chi10,Ho04a}. 
Dramatic progress  in cold-atom experiments make it possible to  achieve  precise measurement of the EOS for two-component fermions in the BEC-BCS crossover \cite{Nas10a,Nas10b,Hor10,Ku12,Hou12}. More recently the EOS for a strongly interacting Bose gas has also been the subject of intensive studies \cite{Rem13,Lau14,Mak14}. However, even in the non-degenerate regime, large three-body atomic losses prevent the achievement of global  thermal equilibrium in a Bose gas \cite{Rem13}. 
In the resonant scattering regime, due to the large separation of scale between the short range of the interatomic forces and the scattering length, few-body systems also have universal properties exemplified by the Efimov effect \cite{Fer10}.
In this context, the virial expansion where the ${n}$th-order term is known from the solution of the $n$-body problem, achieves a remarkable bridge between the few- and the many-body problems \cite{Hua87}. For non degenerate gases and for sufficiently low densities and high-temperatures, the fugacity is a small parameter and the virial expansion is thus a way to derive an accurate EOS. 
Consequently, virial expansion is a subject of  current interest in studies of the two-component Fermi gas \cite{Ho04b,Liu09,Liu10a,Liu10b,Liu10c,Ley11,Dai12,Gha12,Rak12,Liu13,Gao15,Nga15} and also of a Bose gas, where the Efimov effect plays an important role \cite{Cas13,Bar15}. Concerning Bose gases, the momentum distribution has been measured recently \cite{Mak14} and a law at second order of the fugacity including both elastic and inelastic processes has been derived \cite{Lau14}. Based on single-channel models, there are now very accurate evaluations of the third and even of the fourth virial coefficients \cite{Yan16} in homogeneous strongly interacting gases. Currently there are three methods to derive the virial coefficients from a given model: the harmonic oscillator method \cite{Liu10a,Liu10b,Liu10c,Dai12,Rak12,Gha12,Liu13,Cas13,Gao15}, the diagrammatic method  \cite{Ley11}, and the $T$-matrix method \cite{Rei66,Rak12}. At second order of the expansion, Beth and Uhlenbeck showed that the contribution of the interaction to the virial coefficient can be expressed in terms of the scattering phase-shifts for all partial waves \cite{Bet37}. In the context of cold atoms, collisions are dominated by two-body $s$-wave scattering and the second virial coefficient can thus be expressed as a function of the $s$-wave scattering phase-shift.
In current experiments it is also possible to trap cold atoms in highly anisotropic external harmonic potential. This permits one to achieve one-dimensional  (1D) or two-dimensional (2D) systems in the limit where all the characteristic energies, including the temperature, are lower than the atomic zero-point energy of the tight direction(s) of the trap \cite{QGLD}. In this way the 2D EOS for a two-component resonant Fermi gas has been measured \cite{Fen16,Boe16}. On the theoretical side, highly anisotropic traps can be modeled by considering harmonic atomic waveguides where  interatomic collisions are well known in the low energy limit \cite{Ols98,Pet00,Pet01}. The first terms of the virial expansions for strictly 1D or 2D gases using the single-channel approach  have also been studied and  the Beth-Uhlenbeck formula in pure 2D or 1D systems is well established \cite{Cui12,Nga13}.

In this paper, we aim at answering two issues: first, how the molecular state contributes to the Beth-Uhlenbeck formula near a Feshbach resonance and second, how this formula can be generalized when one considers atomic waveguides where several transverse modes are populated and thus, when the dimensionality of the system is between the strict 1D or 2D and the three dimensional limit. The latter issue is especially relevant for atomic waveguides, where few transversal mode are populated. One can wonder how the low-dimensional limit is reached in a regime where the local density approximation (LDA) does not apply.

In this paper, we extend the Beth-Uhlenbeck relation for the second-order virial term to the case of atoms coherently coupled with diatomic molecules in a 2D or a 1D harmonic waveguide. To this end, we use a two-channel modeling of the Feshbach resonance. Our main results are as follows: first, we define the notion of scattering phase-shift in an atomic waveguide in Eq.~\eqref{eq:def_delta}; second, we calculate the second order virial coefficient in Eq.~\eqref{eq:Deltab11_v6} by using the $T$ matrix and the diagrammatic approaches. 

The structure of the paper is as follows. In Sec.~\ref{sec:notations} and Sec.~\ref{sec:equilibrium} we introduce the notations for atoms and molecules in atomic waveguides and consider the thermodynamic equilibrium. This permits us to define the  cumulant, cluster, and virial expansions in Sec.~\ref{sec:cumulants}, Sec.~\ref{sec:cluster} and Sec.~\ref{sec:virial}. The link between the two expansions is done with the cluster expansion in Sec.~\ref{sec:cluster}. By the way, in Appendix we consider the non-interacting gas, which makes it possible to define the domain of validity of the LDA in the thermodynamic limit. In Sec.~\ref{sec:model} we introduce a separable two-channel model already used for bosons or fermions in the vicinity of a magnetic Feshbach resonance  \cite{Wer09,Jon10,Mor11a,Pri11c,Kri15,Tre12}. We then obtain the full $T$ matrix of the model in Sec.~\ref{sec:T_operator} and define the  notion of scattering phase-shift in a waveguide in Sec.~\ref{sec:phase-shift}. In  Sec.~\ref{sec:T-matrix-formalism} we use the $T$ matrix formalism to deduce the Beth-Uhlenbeck relation. Eventually in Sec.~\ref{sec:diagrammatic}, we adapt the method of Ref.~\cite{Ley11} to the two-channel model and deduce the Beth-Uhlenbeck formula from modified Feynman diagrams.

\section{Virial and cumulant expansions in atomic waveguides}

\subsection{Harmonic waveguides}
\label{sec:notations}
In this paper we consider structureless particles (atoms or molecules) in one-dimensional (${D=1}$) or two-dimensional (${D=2}$) waveguides.  Our formalism includes also the case where there is no external potential (${D=3}$). For ${D=2}$, the waveguide is a harmonic oscillator in the ${z}$ direction and for ${D=1}$, it is an isotropic harmonic oscillator in the ${x-y}$ plane. The molecules are created from pairs of atoms by means of the Feshbach mechanism. We denote the trap frequency of the harmonic waveguides ${{\omega}}$ and we assume that it is the same for atoms and molecules considered in this paper  (see the discussion in Appendix B of  Ref.~\cite{Wer09}). The threshold of the energy continuum in the open channel is denoted $E_0$:
\begin{equation}
E_0=\frac{(3-D) \hbar {\omega}}{2} .
\label{eq:E_continuum}
\end{equation}
The internal quantum numbers of the species are labeled by $\eta$: ${\eta={\rm b}}$ for atomic bosons in the open channel, ${\eta=\uparrow, \downarrow}$ for two-component fermions in the open channel, and ${\eta={\rm m}}$ for diatomic molecules in the closed channel. The mass of species ${\eta}$  is then denoted ${m_\eta}$. For convenience, we denote also the mass of the bosonic and $\uparrow$ fermionic species ${m=m_{\rm b}=m_\uparrow}$. The quantum numbers for single particles are denoted ${\alpha}$ and the one-body energy for the  species ${\eta=b,\uparrow,\downarrow}$ is denoted ${\epsilon_\eta^\alpha}$. For example, in 2D waveguides ${\alpha= (k_x,k_y,n_z)}$, where ${(k_x,k_y)}$ are the wavenumbers for the free directions ${(x,y)}$ and ${n_z}$ is the quantum number of the harmonic oscillator associated with the trap along ${z}$ and 
\begin{equation}
\epsilon_\eta^\alpha= \frac{\hbar^2}{2m_\eta} (k_x^2+k_y^2) + \hbar {\omega}
\left( n_z + \frac{1}{2} \right)
\end{equation}

We also introduce, for convenience, the index ${\eta={\rm r}}$  to denote the relative particle associated with two bosons in a Bose gas (or of a pair ${\uparrow \downarrow}$ in a Fermi gas) with  relative mass ${m_{\rm r}=m/2}$ (or ${m_{\rm r}=\frac{m_\downarrow}{1+ m_\downarrow/m}}$). The center of mass of this pair is denoted by the index ${\eta={\rm c}}$ with the mass ${m_{\rm c}=2m}$ for bosons
and ${m_{\rm c}=m+m_\downarrow}$ for fermions. The molecular state has an internal energy  with respect to the 3D open channel continuum denoted ${E_{\rm m}}$. The one-body energy of the molecular state is thus 
\begin{equation}
\epsilon_{\rm m}^\alpha=\epsilon_{\rm c}^\alpha+E_{\rm m}
\end{equation}

\subsection{Thermodynamics at equilibrium} 
\label{sec:equilibrium}
We adopt a grand canonical description of the system at equilibrium at temperature ${T}$ with the grand potentials denoted ${\Omega_{\rm B}}$ (${\Omega_{\rm F}}$) for a Bose gas (for a Fermi gas). The chemical potential for each species ${\eta}$ is denoted ${\mu_\eta}$ and the fugacity is denoted ${z_\eta=e^{\beta \mu_\eta}}$, where ${\beta=1/(k_{\rm B}T)}$. To perform the thermodynamic limit, we assume periodic boundary conditions in the free direction(s) of the atomic waveguides by introducing a $D$-dimensional box of length $L$.  We then consider arbitrarily large values of ${L}$ for a fixed temperature and chemical potentials. We then assume that the summation over a wavenumber, say along the direction $z$ is continuous:
${\sum_{k_z} \equiv  L \int \frac{dk_z}{(2\pi)}}$. The thermodynamic equilibrium between molecules and atoms implies, in a Bose gas,  
\begin{equation}
\mu_{\rm m} = 2 \mu_{\rm b} \quad ; \quad z_{\rm m} =(z_{\rm b})^2
\label{eq:equilibrium_B}
\end{equation}
and in a Fermi gas
\begin{equation} 
\mu_{\rm m} = \mu_\uparrow + \mu_\downarrow \quad ; \quad z_{\rm m} = z_\uparrow z_\downarrow .
\label{eq:equilibrium_F}
\end{equation} 
Consequently, the mean conserved numbers of particles are obtained from the grand potentials by
\begin{equation}
\langle \hat{N}_{\rm b} \rangle +2 \langle  \hat{N}_{\rm m} \rangle= - \frac{\partial \Omega_{\rm B}}{\partial \mu_{\rm b}} \ ; \ 
\langle \hat{N}_{\uparrow/\downarrow} \rangle+\langle  \hat{N}_{\rm m}\rangle= - \frac{\partial \Omega_{\rm F}}{\partial \mu_{\uparrow/\downarrow}} .
\label{eq:mean_conserved_numbers}
\end{equation} 
Another important thermodynamic quantity is the $D$-dimensional spatial density of species ${\eta}$ defined by
\begin{equation}
\rho^\eta_D  = \frac{\langle \hat{N}_\eta \rangle}{L^D}  .
\label{eq:def_rho_D}
\end{equation}
For a 2D (1D) atomic waveguide ${\rho^\eta_D}$ is the areal (the linear) density of species $\eta$. In what follows, we use the one-body canonical partition function in the atomic $D$-dimensional waveguide for the species ${\eta=b}$ and ${\uparrow}$,
\begin{equation}
Q =  \frac{\left( L/\lambda\right)^{{D}} }{\left[2\sinh\left({\beta\hbar{\omega}}/{2}\right) \right]^{3-D}}
\label{eq:Z_one_body}
\end{equation}
where ${\lambda}$ is the de Broglie wavelength 
\begin{equation}
\lambda = \sqrt{\frac{2\pi \beta \hbar^2}{m}}.
\label{eq:lambda_dB}
\end{equation}

\subsection{Cumulant expansion}
\label{sec:cumulants}
Instead of considering directly the expansion of the grand potential in terms of the densities, it is easier to deal with the expansion in terms of the fugacities. Both expansions are sometimes called virial expansion, however we make a distinction between the two by denoting the second one as the cumulant expansion \cite{Cas13}. Following Ref.~ \cite{Dai12}, we define the cumulants ${b_n}$ (${b_{n,p}}$) for a Bose gas (a Fermi gas) with
\begin{equation}
\Omega_{\rm B}=-\frac{Q}{\beta}\sum_{n\ge 1} b_n (z_{\rm b})^n ;
\Omega_{\rm F}=-\frac{Q}{\beta}\sum_{\substack{n,p\ge 0 \\ n+p\ne0}} b_{n,p} (z_{\uparrow})^{n} (z_{\downarrow})^{p}.
\label{eq:def_cumulants}
\end{equation}
The cumulants can thus be deduced from the mean conserved numbers of particles by using Eq.~\eqref{eq:mean_conserved_numbers}. For a Bose gas one obtains
\begin{equation}
\langle \hat{N}_{\rm b} \rangle + 2 \langle \hat{N}_{\rm m}\rangle =Q \sum_{n\ge 1} n \, b_n (z_{\rm b})^n 
\label{eq:Nb_plus_Nm}
\end{equation}
and for a Fermi gas
\begin{align}
\langle \hat{N}_\uparrow\rangle +\langle \hat{N}_{\rm m}\rangle&=Q \sum_{n,p\ge 0} n \, b_{n,p} (z_{\uparrow})^n (z_{\downarrow})^{p}
\label{eq:Nup_plus_Nm} \\
\langle \hat{N}_\downarrow\rangle+\langle \hat{N}_{\rm m}\rangle&=Q\sum_{n,p\ge 0} p\, b_{n,p} (z_{\uparrow})^n (z_{\downarrow})^{p} 
\label{eq:Ndo_plus_Nm} .
\end{align}

\subsection{Cluster expansion}

\label{sec:cluster}

In a Bose gas we define a cluster of order $n$ as the set of eigenstates of the many-body Hamiltonian composed of ${N_{\rm b}}$ bosons and ${N_{\rm m}}$ molecules such that ${N_{\rm b}+2N_{\rm m}=n}$. The canonical partition function for this cluster is denoted ${Q_n}$. In a Fermi gas, a cluster of order ${(n,p)}$ is the set of eigenstates of the many-body Hamiltonian composed of ${N_\uparrow}$ fermions $\uparrow$, ${N_\downarrow}$ fermions $\downarrow$, and ${N_{\rm m}}$ molecules such that ${N_\uparrow+N_{\rm m}=n}$ and ${N_\downarrow+N_{\rm m}=p}$. The canonical partition function for this cluster is denoted ${Q_{n,p}}$. Using Eqs.~\eqref{eq:equilibrium_B} and \eqref{eq:equilibrium_F}, the cluster expansion for the grand potentials are  
\begin{align}
&\Omega_{\rm B} = \frac{-1}{\beta} \ln\left(1+\sum_{n\ge 1} Q_n (z_{\rm b})^n \right) 
\label{eq:clusters_B}
\\
&\Omega_{\rm F} = \frac{-1}{\beta} \ln\left(1+\sum_{\substack{n,p\ge 0 \\ n+p\ne0}} Q_{n,p} (z_{\uparrow})^n (z_{\downarrow})^{p}\right)
\label{eq:clusters_F}
\end{align}
 for small fugacities:
\begin{align}
\Omega_{\rm B}  &=\frac{-1}{\beta} \biggl\{ Q_1 z_{\rm b}+\left[Q_2-\frac{\left(Q_1\right)^2}{2} \right] (z_{\rm b})^2 +\dots\biggr\}
\label{eq:clusters_Omega_B}\\
\Omega_{\rm F}  &= \frac{-1}{\beta} \biggl\{  Q_{1,0} z_\uparrow+Q_{0,1} z_\downarrow+
\left(Q_{1,1}-Q_{1,0}Q_{0,1} \right)z_\uparrow z_\downarrow
 \nonumber\\&
+ \left[Q_{2,0}-\frac{\left(Q_{1,0}\right)^2}{2} \right]
z_\uparrow^2  + \left[Q_{0,2}-\frac{\left(Q_{0,1}\right)^2}{2} \right]z_\downarrow^2 
+  \dots \biggr\}
\label{eq:clusters_Omega_F}
\end{align}
We have ${Q_{1,0}=Q_1=Q}$. The cumulants of Eq.~\eqref{eq:def_cumulants} can thus be expressed in terms of the canonical partition function for a Bose gas,
\begin{equation}
{b_1=1} \quad ; \quad b_2=\frac{Q_2}{Q}-\frac{Q}{2}
\label{eq:clusters_cumulants_B}
\end{equation}
and for a Fermi gas
\begin{align}
&b_{1,0}=1 \ ; \ {b_{0,1}=\frac{Q_{0,1}}{Q}} \ ; \ b_{2,0}=\frac{Q_{2,0}}{Q}-\frac{Q}{2} 
\label{eq:clusters_cumulants_F1}
\\
&b_{0,2}=\frac{Q_{0,2}-\left(Q_{0,1}\right)^2/2}{Q} \quad ; \quad b_{1,1}=\frac{Q_{1,1}}{Q} - Q_{0,1}.
\label{eq:clusters_cumulants_F2}
\end{align}

The expression of the cumulants for a Bose (Fermi) gas without interaction denoted ${b_{n}^{(0)}}$ (${b_{n,p}^{(0)}}$), are given in the Appendix.

\subsection{Virial expansion}
\label{sec:virial}

As a consequence of the interchannel coupling, the number of atoms or molecules are not conserved quantities. Thus, as the virial expansion is a polynomial in terms of densities, we define it in this context by taking into account the conservation laws for ${\hat{N}_{\rm b}+2\hat{N}_{\rm m}}$ and ${\hat{N}_{\uparrow/\downarrow}+\hat{N}_{\rm m}}$. Moreover, due to the presence of a $D$-dimensional atomic waveguide we perform the virial expansion in terms of the densities $\rho^\eta_{D}$ in Eq.~\eqref{eq:def_rho_D}:
\begin{align}
&\Omega_{\rm B}=-\frac{Q}{\beta} \sum_{n\ge 1} \left( \lambda \right)^{nD} a_n \left(\rho^{\rm b}_D + 2 \rho^{\rm m}_D\right)^n 
\label{eq:def_virial_B} \\
&\Omega_{\rm F}=-\frac{Q}{\beta} \sum_{\substack{n,p\ge 0 \\ n+p\ne0}} \left( \lambda \right)^{(n+p)D}  a_{n,p}
\left(\rho^\uparrow_D + \rho^{\rm m}_D \right)^n \nonumber\\
&\qquad \qquad\qquad\qquad \times \left(\rho^\downarrow_D + \rho^{\rm m}_D \right)^{p} .
\label{eq:def_virial_F}
\end{align}
In Eq.~\eqref{eq:def_virial_B} [Eq.~\eqref{eq:def_virial_F}], ${a_n}$ (${a_{n,p}}$) are the virial coefficients for a Bose gas (for a Fermi gas). In this expression the ratio ${\rho^{\rm m}_D/\rho^\eta_D}$ for ${\eta \in \{{\rm b},\uparrow, \downarrow \} }$ is an important quantity which does not appear in the usual one-channel models. It can be evaluated for a two-channel model by using the Hellman-Feynmann theorem \cite{Wer09}. The general relation of the molecular density in terms of the contact permits one to conclude that for a broad resonance ${\rho^{\rm m}_D}$ is also negliglible with respect to the particle density(ies) in the open channel \cite{Pri13}. In the latter regime, one thus recovers the usual virial expansion. 

Combining virial and cumulant expansion of the grand potential together with the relation between the $D$-dimensional density and the cumulants from Eqs.~\eqref{eq:def_rho_D}, \eqref{eq:Nb_plus_Nm}, \eqref{eq:Nup_plus_Nm} and \eqref{eq:Ndo_plus_Nm}, one finds the relation between the virial coefficients and the cumulants : 
\begin{align}
&a_1=\frac{L^D}{\lambda^D Q} \ ; \ a_2=- \frac{L^{2D} b_2}{\lambda^{2D}Q^2}   \\
&a_{1,0}=a_{0,1}=\frac{L^D}{\lambda^D Q} \ ;\  a_{1,1}=\frac{-L^{2D} b_{1,1}}{\lambda^{2D} Q^2 b_{0,1}} \\  
&a_{2,0}=-\frac{L^{2D} b_{2,0}}{\lambda^{2D} Q^2} \  ; \ a_{0,2}=-\frac{L^{2D} b_{0,2}}{\lambda^{2D} Q^2 \left(b_{0,1} \right)^2}.
\end{align}

\section{Beth-Uhlenbeck formula in atomic waveguides}

\subsection{Separable two-channel model}

\label{sec:model}

The atom-atom interaction and the atom-molecule coupling are described by using the separable two-channel model introduced for a two-component Fermi gas in Ref.~\cite{Wer09} and for identical bosons in  Refs.~\cite{Jon10,Mor11a,Pri11c,Tre12,Kri15}. Atoms evolve in the open channel and molecules in the closed channel.  In this simplified model, atoms and molecules are structureless particles and we consider only a single diatomic molecular state. The Feshbach mechanism corresponds to the coherent coupling between  the molecular state and a pair of atoms of reduced mass ${m_{\rm r}}$ and total mass ${m_{\rm c}}$. We denote the free Hamiltonian for a pair of atoms ${\hat{H}^{\rm a}_0}$ and the free Hamiltonian for a single molecule ${\hat{H}^{\rm m}_0}$.  The Hamiltonians ${\hat{H}^{\rm a}_0}$ and ${\hat{H}^{\rm m}_0}$ include the external potential if it exists. For convenience, we adopt a matrix  formalism. The free Hamiltonian for the two-body problem is
\begin{equation}
[\hat{H}_0] = 
\begin{bmatrix}
\hat{H}^{\rm a}_0 & 0\\
0 & \hat{H}^{\rm m}_0
\end{bmatrix}
.
\end{equation}
In this paper, we assume separability between the relative and the center of mass motions. This is the case for the harmonic waveguides considered in this paper. The quantum numbers of the eigenstates of ${\hat{H}^{\rm a}_0}$ are denoted ${(\alpha_{\rm c})}$  for the center of mass and ${(\alpha_{\rm r})}$ for the relative particle. They are denoted ${(\alpha_{\rm m})}$ for the eigenstates of ${\hat{H}^{\rm m}_0}$:
\begin{align}
[\hat{H}_0] 
\begin{bmatrix}
|\alpha_{\rm r} , \alpha_{\rm c} \rangle\\
|\alpha_{\rm m} \rangle
\end{bmatrix}
=
\begin{bmatrix}
 \left( \epsilon_{\rm c}^{\alpha_{\rm c}} + \epsilon_{\rm r}^{\alpha_{\rm r}} \right)  |\alpha_{\rm r} , \alpha_{\rm c} \rangle\\
\epsilon_{\rm m}^{\alpha_{\rm m}} |\alpha_{\rm m} \rangle 
\end{bmatrix}
.
\end{align}
To model the direct interaction and the Feshbach coupling, we introduce the operator 
\begin{equation}
\hat{A}_\epsilon = \hat{\mathbb 1}_{\rm Com} \otimes \langle \delta_\epsilon| 
\label{eq:hat_B}
\end{equation}
where ${\hat{\mathbb 1}_{\rm Com}}$ is the identity operator in the Hilbert space associated with the center of mass of a pair of atoms or of a molecule and the bra ${\langle \delta_\epsilon|}$ acts on the states associated with the relative particle of a pair. In Eq.~\eqref{eq:hat_B} the state ${|\delta_\epsilon\rangle}$ belongs to the Hilbert space of the relative motion for two particles in the open channel and permits one to introduce a cut-off in the model.  For convenience we choose a Gaussian shape and in configuration space and  the momentum representation one has
\begin{equation}
\langle \mathbf r |\delta_\epsilon \rangle = \frac{\exp (-r^2/\epsilon^2)}{\epsilon^3 \pi^{3/2}} \quad ; \quad \langle \mathbf k |\delta_\epsilon \rangle = \exp (-k^2\epsilon^2/4) .
\label{eq:delta_epsilon}
\end{equation}
In Eq.~\eqref{eq:delta_epsilon}, ${\epsilon}$ is the short range parameter of the model. It is of the order of the characteristic radius of the actual potential experienced by atoms or of the size of the actual molecular states  (i.e. of the order of few nanometers).  Using these notations, the direct interaction between two atoms is given by a separable pairwise potential characterized by the strength ${g}$:
\begin{equation}
\hat{V}^{\rm a} = g  \hat{A}_\epsilon^\dagger  \hat{A}_\epsilon .
\label{eq:Vdirect}
\end{equation}
The coherent coupling between a pair of atoms in the open channel and a molecule is 
\begin{equation}
\hat{V}^{\rm ma} = \Lambda   \hat{A}_\epsilon  
\label{eq:Vcoupling}
\end{equation}
where $\Lambda$ is a real parameter. Previous studies showed that using the same state ${|\delta_\epsilon \rangle}$ in the direct  interaction and in the inter-channel coupling  is sufficient to obtain a quantitative description of two-body scattering and shallow bound states \cite{Jon10}. The Hamiltonian of the two-body problem is denoted
\begin{equation}
[\hat{H}]= [\hat{H}_0] + [\hat{V}] ,
\end{equation}
where we have used a matrix notation for the potential of the two-channel model: 
\begin{equation}
[\hat{V}]=
\begin{bmatrix}
\hat{V}^{\rm a} & \hat{V}^{\rm am}\\
\hat{V}^{\rm ma} & 0
\end{bmatrix}
\end{equation}
and ${\hat{V}^{\rm am}=(\hat{V}^{\rm ma})^\dagger}$.

\subsection{Transition operator for the two-channel model}

\label{sec:T_operator}
In what follows, we obtain the transition operator of the model by using the resolvent method. For this purpose we introduce the resolvent for a non-interacting pair of atoms (denoted ${\hat{G}^{\rm a}_0}$) and the resolvent for a molecule (denoted ${\hat{G}^{\rm m}_0}$):
\begin{equation}
\hat{G}^{\rm a}_0({s})=\frac{1}{{s}-\hat{H}^{\rm a}_0} \quad ; \quad \hat{G}^{\rm m}_0({s})=\frac{1}{{s}-\hat{H}^{\rm m}_0} .
\end{equation}
The resolvent of the free two-body Hamiltonian is
\begin{equation}
[\hat{G}_0({s})] = 
\begin{bmatrix}
\hat{G}^{\rm a}_0({s}) & 0\\
0 & \hat{G}^{\rm m}_0({s})
\end{bmatrix} .
\end{equation}
The transition operator has a matrix form
\begin{equation}
[\hat{T}({s})]=
\begin{bmatrix}
\hat{T}^{\rm a}({s}) & \left(\hat{T}^{\rm ma}({s})\right)^\dagger\\
\hat{T}^{\rm ma}({s}) & \hat{T}^{\rm m}({s})
\end{bmatrix}
\end{equation}
and verifies the Lippmann-Schwinger equation
\begin{equation}
[\hat{T}({s})]=[\hat{V}] + [\hat{V}][\hat{G}_0({s})][\hat{T}({s})] .
\label{eq:Lippmann}
\end{equation}
One finds the following matrix elements for the transition operator:
\begin{align}
&\hat{T}^{\rm a}({s})= \hat{V}^{\rm a} + \hat{V}^{\rm am} \hat{G}_0^{\rm m}({s}) \hat{V}^{\rm ma} 
+ \hat{V}^{\rm a} \hat{G}_0^{\rm a}({s}) \hat{T}^{\rm a}({s})   
\nonumber \\
&\qquad \qquad + \hat{V}^{\rm am} \hat{G}_0^{\rm m}({s}) \hat{V}^{\rm ma} \hat{G}_0^{\rm a}({s})  \hat{T}^{\rm a}({s}) .
\label{eq:Lippmann_Ta}\\
&\hat{T}^{\rm ma}({s}) = \hat{V}^{\rm ma} +\hat{V}^{\rm ma} \hat{G}^{\rm a}_0({s}) \hat{T}^{\rm a}({s})\\
&\hat{T}^{\rm m}({s}) = \hat{V}^{\rm ma} \left[ \hat{G}^{\rm a}_0({s})  + \hat{G}^{\rm a}_0({s}) \hat{T}^{\rm a}({s}) \hat{G}^{\rm a}_0({s})\right] \hat{V}^{\rm am} .
\label{eq:Lippmann_Tmol}
\end{align}
Using the separable form of the potentials in Eqs.~\eqref{eq:Vdirect} and \eqref{eq:Vcoupling} one finds
\begin{align}
&\hat{T}^{\rm a}({s}) =\frac{\hat{A}_\epsilon^\dagger  \hat{A}_\epsilon}{ \displaystyle \frac{1}{\hat{g}^{\rm eff}({s})} - \hat{A}_\epsilon \hat{G}^{\rm a}_0({s})\hat{A}_\epsilon^\dagger }
\label{eq:Ta}\\
&\hat{T}^{\rm ma}({s}) = \frac{\hat{V}^{\rm ma}}
{1 - \hat{g}^{\rm eff}({s}) \hat{A}_\epsilon \hat{G}^{\rm a}_0({s})\hat{A}_\epsilon^\dagger}\\
&\hat{T}^{\rm m}({s}) = 
\frac{ |{\Lambda}|^2 \hat{A}_\epsilon \hat{G}^{\rm a}_0({s})\hat{A}_\epsilon^\dagger}
{ 1 - \hat{g}^{\rm eff}({s}) \hat{A}_\epsilon \hat{G}^{\rm a}_0({s}) \hat{A}_\epsilon^\dagger} ,
\label{eq:Tm}
\end{align}
where we have introduced the operator  
\begin{equation}
\hat{g}^{\rm eff}({s}) = g + |\Lambda|^2 \hat{G}^{\rm m}_0({s}) .
\label{eq:operator_geff}
\end{equation}
The eigenvalues of the operator $\hat{g}^{\rm eff}({s})$ in Eq.~\eqref{eq:operator_geff} are given by
\begin{align}
&\hat{g}^{\rm eff}({s}) |\alpha_{\rm c}\rangle = g^{\rm eff}\left({s}^{\rm rel}_{\alpha_{\rm c}}\right)  |\alpha_{\rm c}\rangle \\
&{s}^{\rm rel}_{\alpha_{\rm c}}={s}-\epsilon_{\rm c}^{\alpha_{\rm c}}\\
&g^{\rm eff}({s}^{\rm rel}_{\alpha_{\rm c}}) = g + \frac{|\Lambda|^2}{{s}^{\rm rel}_{\alpha_{\rm c}}-E_{\rm m}} .
\end{align}
Due to the separability, one can define the $T$ matrix of the relative degree of freedom: 
\begin{equation}
\langle \alpha_{\rm r}', \alpha_{\rm c}' | \hat{T}^{\rm a}({s})|\alpha_{\rm r}, \alpha_{\rm c} \rangle =  \langle \alpha_{\rm r}' | \hat{T}^{\rm rel}\left({s}^{\rm rel}_{\alpha_{\rm c}}\right)|\alpha_{\rm r} \rangle \langle \alpha_{\rm c}' | \alpha_{\rm c} \rangle .
\label{eq:Ta_separability}
\end{equation}
One then  recovers the expression of the $T$ matrix used in Ref.~\cite{Kri15}:
\begin{equation}
\langle \alpha_{\rm r}' | \hat{T}^{\rm rel}({s}^{\rm rel}_{\alpha_{\rm c}})|\alpha_{\rm r} \rangle = \frac{\langle \alpha_{\rm r}'|\delta_\epsilon\rangle \langle \delta_\epsilon|\alpha_{\rm r}\rangle }{\displaystyle \frac{1}{g^{\rm eff}({s}^{\rm rel}_{\alpha_{\rm c}})}- 
\langle\delta_\epsilon | \hat{G}^{\rm rel}_0({s}^{\rm rel}_{\alpha_{\rm c}}) | \delta_\epsilon\rangle} .
\label{eq:Trel}
\end{equation}
where we have introduced the resolvent  $\hat{G}^{\rm rel}_0$ for the non-interacting relative motion:
\begin{align}
&H_0^{\rm rel}|\alpha_{\rm r} \rangle=\epsilon_{\rm r}^{\alpha_{\rm r}} |\alpha_{\rm r}\rangle\\
&\hat{G}^{\rm rel}_0({s}) =\frac{1}{{s}-\hat{H}_0^{\rm rel}} .
\label{eq:G0rel}
\end{align}
Using these notations, the diagonal part of the $T$ matrix for the molecule can be written as
\begin{equation}
\langle \alpha_{\rm c}' | \hat{T}^{\rm m}({s})|\alpha_{\rm c} \rangle = \frac{|\Lambda|^2 
\langle\delta_\epsilon | \hat{G}^{\rm rel}_0({s}^{\rm rel}_{\alpha_{\rm c}}) | \delta_\epsilon\rangle  \langle \alpha_{\rm c}' | \alpha_{\rm c} \rangle}{1-{g^{\rm eff}({s}^{\rm rel}_{\alpha_{\rm c}})} 
\langle\delta_\epsilon | \hat{G}^{\rm rel}_0({s}^{\rm rel}_{\alpha_{\rm c}}) | \delta_\epsilon\rangle} .
\label{eq:T_mol}
\end{equation}
The expression of the $T$ matrix in Eq.~\eqref{eq:Trel} can be expressed in terms of the scattering parameters such as the scattering length in the presence of a 2D or 1D  atomic waveguide \cite{Kri15}. 

\subsection{Scattering in atomic waveguides}
\label{sec:phase-shift}
\subsubsection{Multimode scattering regime}

We consider the scattering problem for the relative particle of a pair of bosons or of a pair of ${\uparrow}$ and ${\downarrow}$ fermions in a $D$-dimensional atomic waveguide characterized by the length
\begin{equation}
{l}_{\rm ho} = \sqrt{\frac{\hbar}{m_{\rm r} {\omega} }} .
\end{equation}
For 2D waveguides,  the eigenfunctions of the harmonic trap, are denoted ${\phi_n(z)}$: 
\begin{equation}
\phi_n(z) \equiv \langle z | n \rangle = 
\frac{e^{-\frac{z^2}{2{l}_{\rm ho}^2}}}{ \pi^{1/4} \sqrt{2^n n! {l}_{\rm ho}}}  H_n(z/{l}_{\rm ho}) .
\end{equation}
For the 1D waveguides we use in this section, the cylindrical quantum numbers ${\alpha_{\rm r}\equiv(k,n,m)}$, where ${n}$ is the radial wavenumber of the transverse harmonic oscillator and ${m\hbar}$ is the angular momentum along the axis ${z}$. The single-particle energy of the relative particle is thus
\begin{equation}
\epsilon_{\rm r}^{(k,n,m)}=\frac{\hbar^2 k^2}{2 m_{\rm r}} + \hbar {\omega} \left( 2 n + |m| + 1 \right)
\end{equation}
and the eigenfunction of the transverse harmonic oscillator are given by
\begin{multline}
\langle {\boldsymbol  \rho}|n,m\rangle = \frac{1}{{l}_{\rm ho}} \left[\frac{\pi (n+|m|)!}{n!} \right]^{-1/2} \left(\frac{{\rho}}{{l}_{\rm ho}}\right)^{|m|} e^{im\theta}\\
\times e^{-\frac{1}{2}({\rho}/{l}_{\rm ho})^2}L_n^{(|m|)}({\rho}^2/{l}_{\rm ho}^2) ,
\label{eq:fo_OH_2D}
\end{multline}
where  ${L_n^{(\alpha)}}$ is the generalized Laguerre polynomial and ${ {\boldsymbol  \rho} = ({\rho},\theta)}$ are the polar coordinates in the ${x-y}$ plane.

As a consequence of the separable form of the atomic $T$ matrix where ${|\delta_\epsilon\rangle}$ is a $s$-wave state, scattering occurs only for $s$ waves in free space, for even transverse states  in 2D waveguides, and for  transverse states with zero angular momentum ${(m=0)}$ in 1D waveguides. Thus for a given collisional energy $E$, the number of transverse states that can be populated is given by:
\begin{equation}
N_D(E)=\Bigl\lfloor \frac{E}{2\hbar {\omega}}-\frac{1}{2D}  \Bigr\rfloor 
\label{eq:def_ND}
\end{equation}
where $\lfloor \cdot  \rfloor$ gives the integer part. For ${E> \hbar {\omega} (2+1/D)}$, there are thus several accessible transverse modes for the incoming and outgoing waves in a scattering process. 

\paragraph{Two-dimensional scattering from a single mode incoming state.} 

For a given collisional energy, we define the ${N_D(E)}$ positive wavenumbers ${k_p}$ of the asymptotic scattering states  from the energy conservation:
\begin{equation}
E=\frac{\hbar^2 k_p^2}{2 m_{\rm r}} + \hbar {\omega} \left( 2p  + \frac{1}{2}\right) .
\end{equation}
We consider the scattering state ${|\Psi^{\rm 2D}_{p}\rangle}$ where the incoming wave occupies only one transverse mode: it is characterized by the quantum numbers ${\alpha_{\rm r}=(k_{p} \hat{\mathbf e}_x ,2p)}$, with ${p \le N_2(E)}$. In the limit of large inter-particle distances (${{\rho} \gg {l}_{\rm ho}}$) one has \cite{Kri15}:
\begin{multline}
\langle {\boldsymbol  \rho},z|\Psi^{\rm 2D}_{p}\rangle = 
\langle z | 2p \rangle e^{i k_p  \hat{\mathbf e}_x \cdot {\boldsymbol  \rho}}
-  \frac{m_{\rm r}}{\hbar^2} \sum_{p'=0}^{N_2(E)} \phi_{2p'}(z)\\
\times  
\frac{e^{i\left(k_{p'}{\rho} +\frac{\pi}{4}\right)}}{\sqrt{2\pi k_{p'} {\rho}}} 
\langle k_{p'} \hat{\mathbf e}_{\rho} ,2p'| \hat{T}^{\rm rel}(E+i0^+)| k_{p} \hat{\mathbf e}_x , 2p \rangle ,
\label{eq:2D_scatt_large_rho_sur_epsilon}
\end{multline}
where ${\hat{\mathbf e}_{\rho}}$ is the unitary vector ${{\boldsymbol  \rho}/{\rho}}$.

\paragraph{One-dimensional scattering  from a single-mode incoming state.}

For a transverse mode ${(n, m=0)}$, analogously to the 2D case, we define the wavenumber ${k_n}$ from the energy conservation:
\begin{equation}
E=\frac{\hbar^2 k_n^2}{2 m_{\rm r}} + \hbar {\omega} \left( 2 n  + 1 \right) 
\end{equation}
and consider the scattering state ${|\Psi^{\rm 1D}_{n}\rangle}$ where the incoming wave is characterized by the quantum numbers ${\alpha_{\rm r}=(k_{n} \hat{\mathbf e}_z ,n,0)}$, with ${n \le N_1(E)}$. In the limit of large inter-particle distances (${z \gg {l}_{\rm ho}}$), the scattering wave function verifies that:
\begin{multline}
\langle {\boldsymbol  \rho},z|\Psi^{\rm 1D}_{n}\rangle = 
\langle {\boldsymbol  \rho} | n,0\rangle e^{i k_n z} 
-  \frac{i m_{\rm r}}{\hbar^2} \sum_{n'=0}^{N_1(E)}  \frac{e^{ik_{n'}|z|}}{k_{n'}} \\
\times \langle {\boldsymbol  \rho} | n',0\rangle
 \langle k_{n'},n',0|\hat{T}^{\rm rel}(E +i0^+)| k_n, n,0 \rangle .
\label{eq:1D_scatt_large_rho_sur_epsilon}
\end{multline}

\subsubsection{Scattering phase-shift}

For a given value of the dimension ${D}$ of the waveguide and ${N_D(E)>1}$, the outgoing state in Eq.~\eqref{eq:2D_scatt_large_rho_sur_epsilon} or in Eq.~\eqref{eq:1D_scatt_large_rho_sur_epsilon} occupies  several transverse modes whereas the incoming state occupies only one transverse mode. This prevents defining the notion of scattering phase-shift by using this type of scattering state. To circumvent this problem, we now consider the incoming state ${|\Psi^{E}_D \rangle}$ of relative energy ${E}$, such that the probability of occupation in each transverse mode is conserved in the scattering process. Moreover, as the interaction acts only in the ${s}$-wave sector, we consider only states which are rotationally invariant with respect to ${0z}$ in two-dimensional waveguide and are even function of ${z}$ in one-dimensional waveguide. We expand ${|\Psi^{E}_D \rangle}$ over the basis ${|\Psi^{\rm 2D}_p \rangle}$ or ${|\Psi^{\rm 1D}_p \rangle}$. In the limit of large inter-particle distances one has for the 2D waveguide:
\begin{equation}
\langle {\boldsymbol  \rho},z|\Psi^{E}_2 \rangle = \int \frac{d\theta}{2\pi}
\sum_{p=0}^{N_2(E)} \alpha_{p} 
\langle  {\boldsymbol  \rho},z|\Psi^{\rm 2D}_p \rangle 
\label{eq:2D_scatt_state_over_2p}
\end{equation}
and for the 1D waveguide
\begin{equation}
\langle {\boldsymbol  \rho},z|\Psi^{E}_1 \rangle = \sum_{p=0}^{N_1(E)} \frac{\beta_{p}}{2}
\left( \langle  {\boldsymbol  \rho},z|\Psi^{\rm 1D}_p \rangle +\langle  {\boldsymbol  \rho},-z|\Psi^{\rm 1D}_p \rangle\right) .
\label{eq:1D_scatt_state_over_2p}
\end{equation}
We first consider the scattering phase-shift in the 2D waveguide. After projection over ${\langle z |2p\rangle}$, one obtains in the limit for large values of ${\rho}$,
\begin{multline}
\int dz \langle 2p |z \rangle \langle {\boldsymbol  \rho},z|\Psi^{E}_2 \rangle  = \sqrt{2}
\alpha_{p}  \frac{\sin\left(k_p {\rho} +\frac{\pi}{4}\right)}{\sqrt{\pi k_p {\rho}}}-\frac{m_{\rm r}}{\hbar^2} \\
\times \frac{e^{i\left(k_p {\rho} +\frac{\pi}{4}\right)}}{\sqrt{2\pi k_p {\rho}}} \sum_{p'=0}^{N_2(E)} \alpha_{p'}
\langle k_p \hat{\mathbf e}_{\rho},2p |\hat{T}^{\rm rel}(E +i0^+)| k_{p'}\hat{\mathbf e}_x , 2p' \rangle .
\label{eq:scatt_state_over_2p_bis}
\end{multline}
It is now possible to introduce the notion of scattering phase-shift, denoted ${\delta(E)}$, by identifying Eq.~\eqref{eq:scatt_state_over_2p_bis} with the asymptotic form expected in 2D scattering: 
\begin{multline}
\int dz \phi_{2p}(z) \langle {\boldsymbol  \rho} ,z|\Psi^{E}_2 \rangle  = 
\frac{\alpha_{p} \sqrt{2} e^{i\delta(E)}}{\sqrt{\pi k_p {\rho}}} \\
\times \sin\left[k_p  {\rho} +\delta(E) + \frac{\pi}{4}\right] .
\label{eq:identify_delta}
\end{multline}
Remarkably, one finds formally the same expression as in the free space,
\begin{equation}
 \delta(E)  = \arg \left[ \frac{1}{g_{\rm eff}(E)} - \langle \delta_\epsilon |\hat{G}^{\rm rel}_0(E +i0^+)| \delta_\epsilon\rangle \right]
\label{eq:def_delta}
\end{equation}
and 
\begin{equation}
\alpha_{p}= K \left(\frac{1-\eta}{1+\eta}\right)^{p} \phi_{2p}(0) \exp({-k_p^2 \epsilon^2/4}), 
\end{equation}
where ${\eta= {\epsilon^2}/{(2{l}_{\rm ho}^2)}}$ and ${K}$ is an arbitrary multiplicative constant. It is important to note that the trapping frequency of the atomic waveguide enters implicitly in Eq.~\eqref{eq:def_delta} via the Green's function of the relative motion defined by Eq.~\eqref{eq:G0rel}.

The scattering phase-shift in a 1D waveguide is obtained along the same lines as above. One searchs for the coefficients ${\beta_n}$ in Eq.~\eqref{eq:1D_scatt_state_over_2p} such that at large interparticle distances 
\begin{multline}
\int d^2{\rho} \langle n,m=0 | {\boldsymbol  \rho} \rangle 
\langle {\boldsymbol  \rho},z|\Psi^{E}_1 \rangle = \beta_{n}  e^{i\delta(E)} \\
\times \sin\left[k_n  |z|  +\delta(E) + \frac{\pi}{2}\right] .
\end{multline}
One finds the same expression for the phase-shift as in Eq.~\eqref{eq:def_delta} (but with a different resolvent ${\hat{G}^{\rm rel}_0}$) and
\begin{equation}
\beta_n= K \left(\frac{1-\eta}{1+\eta}\right)^{n} \frac{\exp({-k_n^2 \epsilon^2/4})}{k_n} .
\end{equation}

\subsection{Second order virial coefficient}
\label{sec:T-matrix-formalism}

\begin{figure}
\centerline{\includegraphics[width=8cm]{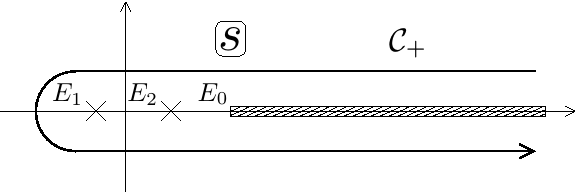}}
\caption{Example of the contour of integration in Eq.~\eqref{eq:Reiner} where ${E_1}$ and ${E_2}$ are the energies of two-bound states.}
\label{fig:contour}
\end{figure}

The second order virial coefficient is related to the resolvent of the two-channel model
\begin{equation}
[\hat{G}({s})] = \frac{1}{{s}-[\hat{H}]}
\end{equation}
through the identity 
\begin{equation}
{\rm Tr} \left\{ e^{\beta [\hat{H}]} \right\} = \int_{\mathcal C_+} \frac{d{s}\, e^{-\beta {s}}}{2i\pi}
 {\rm Tr} \left\{ [\hat{G}({s})] \right\} 
\label{eq:Reiner}
\end{equation}
where ${\mathcal C_+}$ is a  counter clockwise contour from ${\infty+i0^+}$ to ${\infty-i0^+}$ in a loop which encompasses all the poles of the integrand and the branch cut associated with the energy continuum of the energy continuum on the real axis. An example of such a contour is plotted in Fig.~(\ref{fig:contour}). 
Using the relation between the transition operator and the resolvent of the Hamiltonian
\begin{equation}
[\hat{G}({s})] - [\hat{G}_0({s})] = [\hat{G}_0({s})][\hat{T}({s})][\hat{G}_0({s})] ,
\end{equation}
one can find the contribution of the interaction ${\Delta b_{1,1}}$ and ${\Delta b_{2}}$ for the second order virial coefficient. In the case of a two-component Fermi gas, the non interacting contributions ${b_{1,1}^{(0)}}$, ${b_{2,0}^{(0)}}$ and ${b_{0,2}^{(0)}}$ are given in Eqs.~\eqref{eq:cumulants_free_F} and \eqref{eq:b110_3D_F}. The contribution of the interaction is then ${\Delta b_{1,1}= b_{1,1}-b_{1,1}^{(0)}}$ \cite{Rei66,Nga15,no-pwave} with:
\begin{equation}
\Delta b_{1,1} =  \frac{1}{Q} \int_{\mathcal C_+} \frac{d{s}\, e^{-\beta {s}}}{2i\pi}  {\rm Tr} \left\{ [\hat{G}_0({s})][\hat{T}({s})][\hat{G}_0({s})] \right\} .
\label{eq:Deltab11_v1}
\end{equation}
The trace in Eq.~\eqref{eq:Deltab11_v1} can be decomposed in the atomic and molecular sectors:
\begin{multline}
\Delta b_{1,1} = \frac{1}{Q} \int_{\mathcal C_+} \frac{d{s}\, e^{-\beta {s}}}{2i\pi}  {\rm Tr} \left\{\hat{G}^{\rm a}_0({s})\hat{T}^{\rm a}({s}) \hat{G}^{\rm a}_0({s}) \right.\\
\left. + \hat{G}^{\rm m}_0({s})\hat{T}^{\rm m}({s}) \hat{G}^{\rm m}_0({s})
\right\} .
\label{eq:Deltab11_v2}
\end{multline}
The correction to the second virial coefficient is then
\begin{multline}
\Delta b_{1,1} = \frac{1}{Q} \int_{\mathcal C_+} \frac{d{s}\, e^{-\beta {s}}}{2i\pi} 
\left\{
\sum_{\alpha_{\rm r},\alpha_{\rm c}} \frac{\langle \alpha_{\rm r} | \hat{T}^{\rm rel}({s}^{\rm rel}_{\alpha_{\rm c}})|\alpha_{\rm r} \rangle}{\left[{s}^{\rm rel}_{\alpha_{\rm c}}-\mathcal E_{\rm r}(\alpha_{\rm r})\right]^2} \right. \\
\left.
+ \sum_{\alpha_{\rm c}} \frac{\langle \alpha_{\rm c} | \hat{T}^{\rm m}({s})|\alpha_{\rm c} \rangle}{\left({s}^{\rm rel}_{\alpha_{\rm c}}- E_{\rm m}\right)^2} \right\} .
\label{eq:Deltab11_v3}
\end{multline}
It is then fruitful to use the identity
\begin{equation}
\sum_{\alpha_{\rm r}} \frac{|\langle \alpha_{\rm r}|\delta_\epsilon\rangle|^2}{\left[{s}^{\rm rel}_{\alpha_{\rm c}}-\mathcal E_{\rm r}(\alpha_{\rm r})\right]^2} = -\frac{d}{d{s}^{\rm rel}_{\alpha_{\rm c}}} \langle\delta_\epsilon | \hat{G}^{\rm rel}_0({s}^{\rm rel}_{\alpha_{\rm c}}) | \delta_\epsilon\rangle ,
\end{equation}
which permits one to obtain 
\begin{multline}
\frac{\langle \alpha_{\rm c} | \hat{T}^{\rm m}({s})|\alpha_{\rm c} \rangle}{\left({s}^{\rm rel}_{\alpha_{\rm c}}- E_{\rm m}\right)^2} + \sum_{\alpha_{\rm r}} \frac{\langle \alpha_{\rm r} | \hat{T}^{\rm rel}\left({s}^{\rm rel}_{\alpha_{\rm c}}\right)|\alpha_{\rm r} \rangle}{\left[{s}^{\rm rel}_{\alpha_{\rm c}}-\mathcal E_{\rm r}(\alpha_{\rm r})\right]^2}  \\
=\frac{d}{d{s}^{\rm rel}_{\alpha_{\rm c}}} \ln \left[{1-g^{\rm eff}({s}^{\rm rel}_{\alpha_{\rm c}}) 
\langle\delta_\epsilon | \hat{G}^{\rm rel}_0({s}^{\rm rel}_{\alpha_{\rm c}}) | \delta_\epsilon\rangle }\right] .
\end{multline}
By making the change of variable ${{s}={s}^{\rm rel}_{\alpha_{\rm c}}+\epsilon_{\rm c}^{\alpha_{\rm c}}}$ in the integral in Eq.~\eqref{eq:Deltab11_v3}, one can then perform the summation over  ${\alpha_{\rm c}}$ with 
\begin{equation}
\sum_{\alpha_{\rm c}} e^{-\beta \epsilon_{\rm c}^{\alpha_{\rm c}}} = \left(\frac{m_{\rm c}}{m_\uparrow}\right)^{D\over2}Q ,
\end{equation}
which gives  the contribution of the center of mass degree of freedom. Equation~\eqref{eq:Deltab11_v3} can then be expressed as:
\begin{multline}
\Delta b_{1,1} =  \left(\frac{m_{\rm c}}{m_\uparrow}\right)^{D\over2} \int_{\mathcal C_+} \frac{d{s}\, e^{-\beta {s}}}{2i\pi} 
\left\{
\frac{d}{d{s}} \ln \left[g^{\rm eff}({s}) \right]
 \right.\\ 
+ \left. \frac{d}{ds} \ln \left[\frac{1}{g^{\rm eff}({s})} - \langle\delta_\epsilon | \hat{G}^{\rm rel}_0({s}) | \delta_\epsilon\rangle \right] \right\} .
\label{eq:Deltab11_v4}
\end{multline}
One recognizes in the second term of the integral in  Eq.~\eqref{eq:Deltab11_v4} the denominator of the relative $T$ matrix of Eq.~\eqref{eq:Trel}. As shown in Ref.~\cite{Kri15}, for the separable two-channel model used in this paper there are at most two dimers in atomic waveguides. Their energies denoted ${E_1}$, and ${E_2}$ are thus simple poles of this second term. This latter term also has a pole at ${{s}=E_{\rm m}-\frac{|\Lambda|^2}{g}}$. The first term in the integral in Eq.~\eqref{eq:Deltab11_v4} has a pole at ${s}=E_{\rm m}$ and a pole at ${s=E_{\rm m}-\frac{|\Lambda|^2}{g}}$. The two residues at ${s=E_{\rm m}-\frac{|\Lambda|^2}{g}}$ cancel each other and one finds:
\begin{multline}
\Delta b_{1,1} =  \left(\frac{m_{\rm c}}{m_\uparrow}\right)^{D\over2} \left\{ -e^{-\beta E_{\rm m}} + \sum_i e^{-\beta E_i}
-  \int_{E_0}^\infty \frac{d{s}}{2i\pi} \right.\\
\left. \times
e^{-\beta {s}}\frac{d}{d{s}} \ln \left[\frac{\frac{1}{g^{\rm eff}({s})} - \langle\delta_\epsilon | \hat{G}^{\rm rel}_0({s}+i0^+) | \delta_\epsilon\rangle} {\frac{1}{g^{\rm eff}({s})} - \langle\delta_\epsilon | \hat{G}^{\rm rel}_0({s}-i0^+) | \delta_\epsilon\rangle}\right] 
\right\} ,
\label{eq:Deltab11_v5}
\end{multline}
where ${E_0}$ is the threshold for the continuum of the energy spectrum given in Eq.~\eqref{eq:E_continuum}. From Eq.~\eqref{eq:G0rel}, the resolvent verifies:
\begin{equation}
\langle\delta_\epsilon | \hat{G}^{\rm rel}_0({s}-i0^+) | \delta_\epsilon\rangle
= \langle\delta_\epsilon | \hat{G}^{\rm rel}_0({s}+i0^+) | \delta_\epsilon\rangle^*
\end{equation}
and one thus obtains from the expression of the scattering phase-shift in Eq.~\eqref{eq:def_delta}
\begin{equation}
\frac{\frac{1}{g^{\rm eff}({s})} - \langle\delta_\epsilon | \hat{G}^{\rm rel}_0({s}+i0^+) | \delta_\epsilon\rangle} {\frac{1}{g^{\rm eff}({s})} - \langle\delta_\epsilon | \hat{G}^{\rm rel}_0({s}-i0^+) | \delta_\epsilon\rangle}
=e^{-2i\delta({s})} .
\end{equation}
This permits one to extend the Beth-Uhlenbeck formula to the case of harmonic atomic waveguides
\begin{multline}
\Delta b_{1,1} =  \left(1+\frac{m_\downarrow}{m_\uparrow}\right)^{D\over2} \bigg\{ -e^{-\beta E_{\rm m}} + \sum_i e^{-\beta E_i}
\\
+  \int_{E_0}^\infty \frac{d{s}}{\pi} e^{-\beta {s}}\frac{d\delta({s})}{d{s}} 
\bigg\} .
\label{eq:Deltab11_v6}
\end{multline}
Remarkably, the explicit molecular contribution in  Eq.~\eqref{eq:Deltab11_v6} exactly cancels the one in the non-interacting virial coefficient of Eq.~\eqref{eq:b110_3D_F}. For a Bose gas, the correction ${\Delta b_2=b_2-b_2^{(0)}}$ is derived along the same lines as above. The corresponding Beth-Uhlenbeck formula is obtained from Eq.~\eqref{eq:Deltab11_v6} by making the substitution ${m_\uparrow \to m}$ and ${\Delta b_{1,1} \to \Delta b_2}$.

\section{Diagrammatic method in atomic waveguides}
\label{sec:diagrammatic}
\subsection{Green's functions}
\label{sec:green}
An alternative way of obtaining the virial coefficients is to deduce them from the population of the different atomic and molecular species by using Eqs.~\eqref{eq:Nb_plus_Nm}, \eqref{eq:Nup_plus_Nm} and \eqref{eq:Ndo_plus_Nm}. The diagrammatic method introduced in Ref.~\cite{Ley11} permits one to obtain the populations and thus the virial coefficients from a systematic expansion of the Green's function in powers of the fugacities. Here, the diagrammatic method is adapted for the two-channel model in the presence of an atomic waveguide.

We introduce the creation operator ${\hat{a}_{\eta,\alpha}^\dagger}$ for a particle of the species ${\eta\in\{\uparrow,\downarrow,{\rm b},{\rm m}\}}$ with an external quantum number denoted ${\alpha}$. Depending on the statistics of the particles ${\eta}$, ${\hat{a}_{\eta,\alpha}}$ and ${\hat{a}_{\eta,\alpha}^\dagger}$ verify the standard commutation ar anticommutation relations. For convenience, we introduce the variable ${\chi_\eta=\pm 1}$ depending on the statistics of species ${\eta}$:
 \begin{equation}
\chi_{\rm b} = \chi_{\rm m} = 1 \ ; \ \chi_\uparrow=\chi_\downarrow = -1.
\label{eq:statistics}
\end{equation}

\begin{figure}
\centerline{\includegraphics[width=8cm]{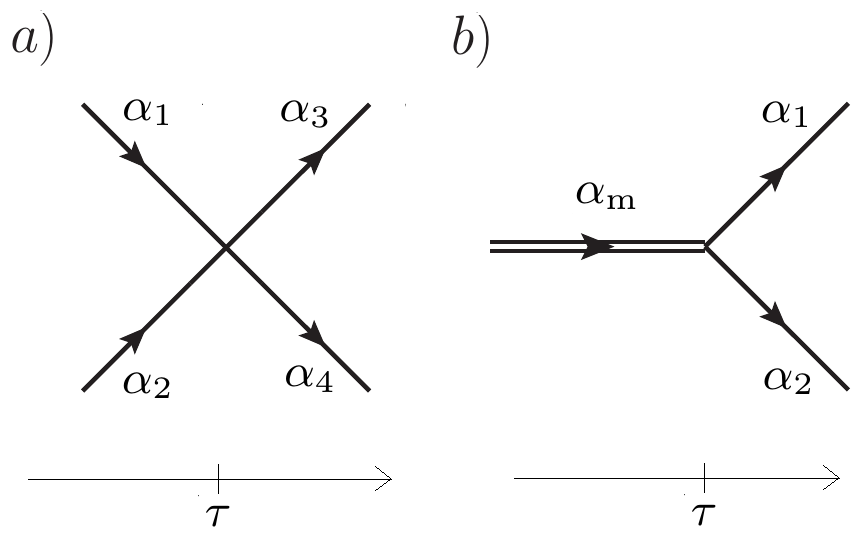}}
\caption{$a)$ Vertex associated with the direct interaction in the finite-temperature formalism :
$g \langle \alpha_1 \alpha_2 | \hat{A}_\epsilon^\dagger \hat{A}_\epsilon |  \alpha_3 \alpha_4 \rangle \delta(\tau)$;
$b)$  idem for the interchannel coupling  :
$\Lambda \langle \alpha_{\rm m}| \hat{A}_\epsilon | \alpha_1 \alpha_2 \rangle \delta(\tau)$.
}
\label{fig:vertex}
\end{figure}

The populations of the different species are related to the Green's functions at finite temperature ${G_\eta(\alpha,\tau)=-\langle T_\tau \hat{a}_{\eta,\alpha}(\tau) \hat{a}_{\eta,\alpha}^\dagger(0) \rangle}$ via the well-known formula ${\langle \hat{N}_\eta\rangle=-\chi_\eta \sum_\alpha G_\eta(\alpha,\tau=0^-)}$. As in standard diagrammatic methods, we then introduce the free Green's function:
\begin{equation}
G^{0}_\eta(\alpha,\tau)=-e^{-(\epsilon_\eta^\alpha-\mu_\eta)\tau/\hbar} \left[ \theta(\tau) + \chi_\eta n_\eta(\epsilon_\eta^\alpha-\mu_\eta) \right]
\label{eq:Green} 
\end{equation}
where ${\theta(\tau)}$ is the Heaviside's function. In the diagrammatic expansions we introduce a solid line for the atomic Green's functions and a double solid line for the molecular Green's function. The bare vertices for the direct interaction in Eq.~\eqref{eq:Vdirect} and for the interchannel coupling in Eq.~\eqref{eq:Vcoupling} are depicted in Fig.~(\ref{fig:vertex}).
Following Ref.~\cite{Ley11}, we perform the expansion of the Green's function in Eq.~\eqref{eq:Green} in terms of the fugacity,
\begin{multline}
G^{0}_\eta(\alpha,\tau) = e^{\mu_\eta \tau/\hbar} \left[G^{(0,0)}_\eta(\alpha,\tau) + G^{(0,1)}_\eta (\alpha,\tau) z_\eta \right.\\ 
\left. +  G^{(0,2)}_\eta(\alpha,\tau) z_\eta^2 + \dots \right].
\label{eq:expansion_G0}
\end{multline}
In Eq.~\eqref{eq:expansion_G0}, ${G^{(0,0)}_\eta(\alpha,\tau)}$ corresponds to the high-temperature limit of the Green's function in vacuum where the fugacity is 0. An important fact for the diagrammatic approach is that it is a purely retarded function :
\begin{equation}
G^{(0,0)}_\eta(\alpha,\tau)  = - \theta(\tau) e^{-\epsilon_\eta^\alpha \tau/\hbar} .
\end{equation}
The other orders in Eq.~\eqref{eq:expansion_G0} are given by
\begin{equation}
 G^{(0,n)}_\eta (\alpha,\tau)  = - (\chi_\eta)^n e^{-\epsilon_\eta^\alpha( n\beta+\tau /\hbar)} .
\end{equation}
We then introduce the  modified Feynman diagrams from Ref.~\cite{Ley11} for the diagrammatic expansions in terms of the fugacities: an $n$-slashed line (or double line) is associated with the functions ${G^{(0,n)}_\eta (\alpha,\tau)}$, and by definition a solid non slashed line corresponds to the function ${G^{(0,0)}_\eta(\alpha,\tau)}$.

\subsection{Two-particle vertex in the high-temperature limit}
\label{sec:vertex}

\begin{figure*}
\includegraphics[width=\textwidth]{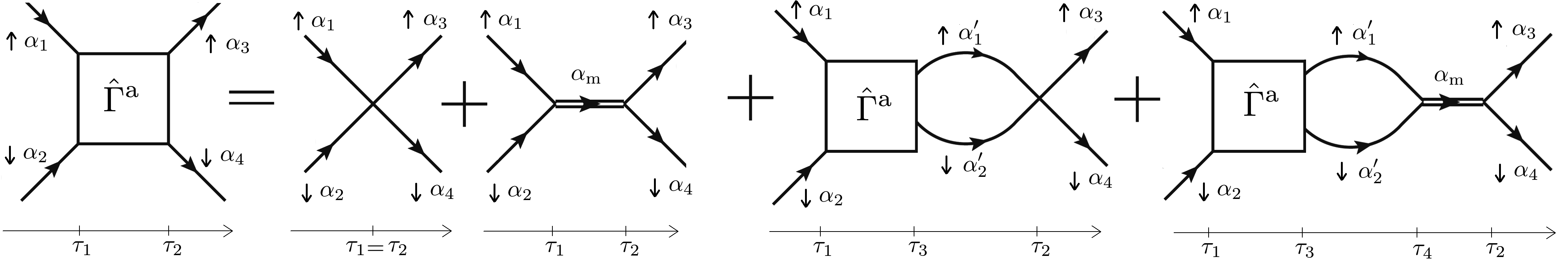}
\caption{Diagrammatic form of the Bethe-Salpeter equation for the atom-atom vertex $\hat{\Gamma}^{\rm a}$.}
\label{fig:Bethe-Salpeter_atoms}
\end{figure*}

In what follows, we show how the two-particle vertex in the high-temperature limit is simply related to the two-body $T$ matrix. We consider the fermionic case. The high-temperature limit corresponds to a low density limit, where two particle collisions are obtained in the ladder approximation. For the two-particle vertex function, we thus consider the  diagrammatic expansion in Fig.~(\ref{fig:Bethe-Salpeter_atoms}). 
The vertex for the atom-atom interaction is then given by the Bethe-Salpeter equation:
\begin{widetext}
\begin{multline}
\langle \alpha_1,\alpha_2|\hat{\Gamma}^{\rm a}(\tau_2-\tau_1)|\alpha_3,\alpha_4\rangle
=\langle \alpha_1,\alpha_2 | \hat{V}^{\rm a} | \alpha_3,\alpha_4 \rangle \delta(\tau_2-\tau_1)
-\frac{1}{\hbar} \sum_{\alpha_{\rm m}} \langle \alpha_1,\alpha_2 |\hat{V}^{\rm am}|\alpha_{\rm m}
\rangle G_{\rm m}^0(\alpha_{\rm m},\tau_2-\tau_1) \langle \alpha_{\rm m} | \hat{V}^{\rm ma}|\alpha_3,\alpha_4 \rangle\\
-\frac{1}{\hbar} \int_0^{\beta \hbar} d\tau_3 \sum_{\alpha_1',\alpha_2'} \langle \alpha_1,\alpha_2|\hat{\Gamma}^{\rm a}(\tau_3-\tau_1)|\alpha_1',\alpha_2'\rangle
G_\uparrow^0(\alpha_1',\tau_2-\tau_3) 
G_\downarrow^0(\alpha_2',\tau_2-\tau_3)
\langle \alpha_1',\alpha_2' | \hat{V}^{\rm a} | \alpha_3,\alpha_4 \rangle\\
+ \frac{1}{\hbar^2} \int_0^{\beta \hbar} d\tau_3 \int_0^{\beta \hbar} d\tau_4 \sum_{\alpha_1',\alpha_2',\alpha_{\rm m}} 
\langle \alpha_1,\alpha_2|\hat{\Gamma}^{\rm a}(\tau_3-\tau_1)|\alpha_1',\alpha_2'\rangle
G_\uparrow^0(\alpha_1',\tau_4-\tau_3) 
G_\downarrow^0(\alpha_2',\tau_4-\tau_3)\\
\times
\langle \alpha_1',\alpha_2' | \hat{V}^{\rm am} | \alpha_{\rm m} \rangle
G_{\rm m}^0(\alpha_{\rm m},\tau_2-\tau_4) \langle \alpha_{\rm m} |  \hat{V}^{\rm ma} | \alpha_3,\alpha_4 \rangle .
\label{eq:Bethe_Salpeter_at}
\end{multline}
\end{widetext}
In the high-temperature limit, the free particle Green's functions  are given by the first term in the expansion of Eq.~\eqref{eq:expansion_G0}. Consequently, in this limit there are only retarded functions and there is no ${\beta}$ dependence in Eq.~\eqref{eq:Bethe_Salpeter_at}. We can thus replace the upper limit of the integration interval for the intermediate times with ${+\infty}$. We also factorize the global time imaginary dependence to extract the vacuum contribution:
\begin{equation}
\lim_{\{z_\eta\} \to 0} \hat{\Gamma}^{\rm a}(\tau_2-\tau_1)  = e^{(\mu_\uparrow+\mu_\downarrow)(\tau_2-\tau_1)}\, \hat{\Gamma}^{\rm a}_{\rm vac}.
\end{equation}
We then use the separability of the two-channel model, which permits one to decouple the center of mass from the relative degree of freedom with identity
\begin{multline}
\sum_{\alpha_1',\alpha_2'} |\alpha_1',\alpha_2'\rangle G^{(0,0)}_\downarrow(\alpha_1',\tau) G^{(0,0)}_\uparrow(\alpha_2',\tau) 
\langle \alpha_1',\alpha_2'|\\
=\sum_{\alpha_{\rm r},\alpha_{\rm c}} |\alpha_{\rm r},\alpha_{\rm c} \rangle \langle \alpha_{\rm r},\alpha_{\rm c} | 
e^{-\hat{H}_0^{\rm a} \tau} \theta(\tau) .
\label{eq:separability}
\end{multline}
The atomic part of the $T$ matrix is  the Laplace transform of the vacuum vertex function:
\begin{align}
&\hat{T}^{\rm a}({s}) = \int_{0}^\infty d\tau e^{{s} \tau/\hbar} \hat{\Gamma}^{\rm a}_{\rm vac}(\tau) 
\label{eq:Laplace_transform}\\
&\hat{\Gamma}^{\rm a}_{\rm vac}(\tau) = \int_{\mathcal C_\gamma} \frac{d{s}}{2i\pi \hbar} e^{ - {s} \tau/\hbar} \hat{T}^{\rm a}({s}) .
\label{eq:inverse_Laplace_transform}
\end{align}
The contour ${\mathcal C_\gamma}$ in the inverse Laplace transform of Eq.~\eqref{eq:inverse_Laplace_transform} is the vertical line in the complex plane ${\Re({s})=\gamma \in \mathbb R}$, oriented from ${\gamma-i\infty}$ to ${\gamma+i\infty}$ and such that the integrand is an analytical function on the left-hand side of the contour. In the high-temperature limit of Eq.~\eqref{eq:Bethe_Salpeter_at}, the functions in the integrands are retarded and the intermediate imaginary times are thus ordered. This permits one to simplify the calculation of the Laplace transform in Eq.~\eqref{eq:Laplace_transform}. For example the contribution of the third term on the right hand-side of Eq.~\eqref{eq:Bethe_Salpeter_at}, where we set ${\tau_1=0}$, can be evaluated by using the following change of variable:
\begin{multline}
\int_0^\infty d\tau_2 e^{{s} \tau_2/\hbar}  \int_0^{\infty} d\tau_3 \hat{\Gamma}^{\rm a}_{\rm vac}(\tau_3) e^{-\hat{H}_0^{\rm a} (\tau_2-\tau_3)/\hbar} \theta(\tau_2-\tau_3)\\
=  \int_0^\infty d\tau e^{({s}-\hat{H}_0^{\rm a}) \tau/\hbar}  \int_0^\infty d\tau_3 e^{{s}\tau_3/\hbar}\hat{\Gamma}^{\rm a}_{\rm vac}(\tau_3) .
\end{multline}
We use the same type of change of variable in the fourth term on the right hand side of Eq.~\eqref{eq:Bethe_Salpeter_at} and eventually we obtain the Lippmann-Schwinger equation \eqref{eq:Lippmann_Ta}.  The expression of the atomic part of the transition matrix in Eq.~\eqref{eq:Ta} follows directly from the latter equation. Note that this result could also have been  obtained in the zero temperature formalism in vacuum. 

\begin{figure*}
\includegraphics[width=\textwidth]{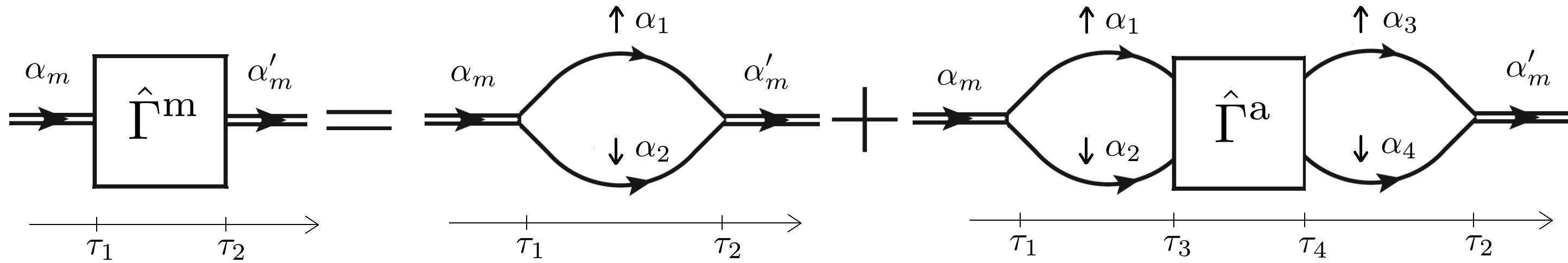}
\caption{Diagrammatic form of  Eq.~\eqref{eq:Bethe_Salpeter_mol}.}
\label{fig:Bethe-Salpeter_mol}
\end{figure*}

Concerning the molecular part, we introduce the operator ${\hat{\Gamma}^{\rm m}}$ analogous to ${\hat{\Gamma}^{\rm a}}$ with the diagrammatic expansion given in Fig.~(\ref{fig:Bethe-Salpeter_mol}). 
We thus obtain the Bethe-Salpeter like equation:
\begin{widetext}
\begin{multline}
\langle \alpha_m|\hat{\Gamma}^{\rm m}(\tau_2-\tau_1)|\alpha_m' \rangle
= \sum_{\alpha_1,\alpha_2} \langle \alpha_{\rm m}| \hat{V}^{\rm ma} | \alpha_1,\alpha_2 \rangle
G_\uparrow^{0}(\alpha_1,\tau_2-\tau_1) 
G_\downarrow^{0}(\alpha_2,\tau_2-\tau_1)
\langle \alpha_1,\alpha_2  |  \hat{V}^{\rm am } | \alpha_{\rm m}' \rangle
\\
- \frac{1}{\hbar^2} \int_0^{\beta\hbar} d\tau_3 \int_0^{\beta\hbar} d\tau_4 \sum_{\alpha_1,\alpha_2,\alpha_3,\alpha_4}
\langle \alpha_{\rm m}| \hat{V}^{\rm ma} | \alpha_1,\alpha_2 \rangle
G_\uparrow^{0}(\alpha_1,\tau_3-\tau_1) 
G_\downarrow^{0}(\alpha_2,\tau_3-\tau_1) \\
\times 
\langle \alpha_1,\alpha_2|\hat{\Gamma}^{\rm a}(\tau_4-\tau_3)|\alpha_3,\alpha_4\rangle
G_\uparrow^{0}(\alpha_3,\tau_2-\tau_4) G_\downarrow^{0}(\alpha_4,\tau_2-\tau_4)
\langle \alpha_3,\alpha_4 |\hat{V}^{\rm am} | \alpha_{\rm m}'\rangle,
\label{eq:Bethe_Salpeter_mol}
\end{multline}
\end{widetext}
In the high-temperature limit, we introduce the vacuum contribution ${\hat{\Gamma}^{\rm m}_{\rm vac}}$  along the same lines as was done for the atomic part and we recover the molecular part of the $T$ matrix in Eq.~\eqref{eq:T_mol} with
\begin{equation}
\hat{T}^{\rm m}({s}) = - \int_{0}^\infty d\tau e^{{s}\tau/\hbar} \hat{\Gamma}^{\rm m}_{\rm vac}(\tau) .
\end{equation}
Equation \eqref{eq:Bethe_Salpeter_mol} can then be transformed into Eq.~\eqref{eq:Lippmann_Tmol}, and eventually one recovers Eq.~\eqref{eq:Tm}.

\subsection{The Beth-Uhlenbeck formula from the diagrammatic method}

\begin{figure}[h]
\centerline{\includegraphics[width=8cm]{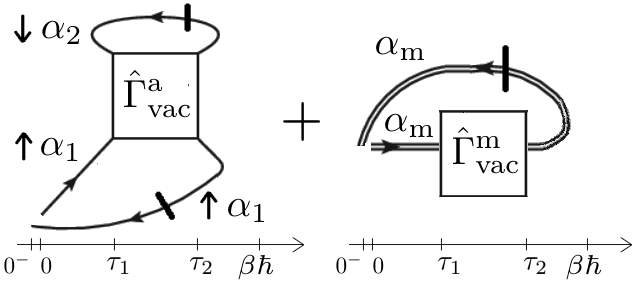}}
\caption{The two diagrams contributing to ${\Delta b_{1,1}}$, see Eqs.~\eqref{eq:defDeltab_11}, \eqref{eq:diag_second_order}.}
\label{fig:diag_Deltab_11}
\end{figure}

Here, we evaluate the second order virial coefficient for the two-spin component Fermi gas by using the diagrammatic method. We denote the second-order term in the expansion of Eq.~\eqref{eq:Nup_plus_Nm} ${\delta^{(2)} \langle \hat{N}_\uparrow\rangle+ \delta^{(2)} \langle \hat{N}_{\rm m}\rangle}$. The second order virial coefficient is then obtained from the identity
\begin{equation}
 \Delta b_{1,1}= \frac{1}{z_\uparrow z_\downarrow Q} \left(\delta^{(2)} \langle \hat{N}_\uparrow\rangle + \delta^{(2)} \langle \hat{N}_{\rm m}\rangle \right) .
 \label{eq:defDeltab_11}
\end{equation}
In Fig.~(\ref{fig:diag_Deltab_11}), we  express Eq.~\eqref{eq:defDeltab_11} in terms of the Feynman and blocks diagrams introduced in Sec.~\ref{sec:green} and Sec.~\ref{sec:vertex} . 
The corresponding equation is:
\begin{multline}
 \Delta b_{1,1}= \frac{1}{\hbar Q} \int_0^{\beta \hbar} d\tau_1 \int_0^{\beta \hbar} d\tau_2
\bigg[ \sum_{\alpha_1,\alpha_2} G_\uparrow^{(0,0)}(\alpha_1,\tau_1) \\
\times G_\downarrow^{(0,1)}(\alpha_2,\tau_1-\tau_2)
G_\uparrow^{(0,1)}(\alpha_1,0^--\tau_2)  \\
\times \langle \alpha_1,\alpha_2 | \hat{\Gamma}^{\rm a}_{\rm vac}(\tau_2-\tau_1) | \alpha_1,\alpha_2 \rangle
- \sum_{\alpha_{\rm m}} G_{\rm m}^{(0,0)}(\alpha_m,\tau_1)  \\
\times G_{\rm m}^{(0,1)}(\alpha_{\rm m},0^--\tau_2)
\langle \alpha_{\rm m} | \hat{\Gamma}^{\rm m}_{\rm vac}(\tau_2-\tau_1) | \alpha_{\rm m} \rangle\bigg] .
\label{eq:diag_second_order}
\end{multline}
By using the same type of simplification as in Eq.~\eqref{eq:separability} and performing the change of variable ${\tau_2=\tau+\tau_1}$ one finds
\begin{multline}
 \Delta b_{1,1} = \frac{-1}{\hbar Q} \int_0^\infty d\tau_1 \int_0^\infty d\tau
 \theta(\beta \hbar-\tau_1-\tau)  \\
\times {\rm Tr} \left[ e^{-\hat{H}_0^{\rm a}(\beta -\tau/\hbar)} \hat{\Gamma}_{\rm vac}^{\rm a}(\tau)
 +  e^{-\hat{H}_0^{\rm m}(\beta - \tau/\hbar)} \hat{\Gamma}^{\rm m}_{\rm vac}(\tau) \right] 
\end{multline}
and the integration over ${\tau_1}$ yields
\begin{multline}
\Delta b_{1,1} =  \frac{-1}{\hbar Q} \int_0^{\infty} d\tau
 (\beta \hbar -\tau) \theta(\beta \hbar-\tau)
\times {\rm Tr} \left[ e^{-\hat{H}_0^{\rm a}(\beta -\tau/\hbar)} 
\right.\\
\left. \times \hat{\Gamma}_{\rm vac}^{\rm a}(\tau) +  e^{-\hat{H}_0^{\rm m}(\beta - \tau/\hbar)} \hat{\Gamma}^{\rm m}_{\rm vac}(\tau) \right] .
\label{eq:convol}
\end{multline}
Equation \eqref{eq:convol} is a product of the convolution of retarded functions evaluated at the imaginary time ${\beta\hbar}$. It can thus be expressed as the inverse Laplace transform of a product of two Laplace transforms:
\begin{equation}
\Delta b_{1,1} =  -  \int_{\mathcal C_\gamma} \frac{d{s}}{2i\pi} \frac{e^{ - \beta {s}}}{Q} 
{\rm Tr} \bigg[\frac{\hat{T}^{\rm a}({s})}{({s}-\hat{H}_0^{\rm a})^2} 
 + \frac{\hat{T}^{\rm m}({s})}{({s}-\hat{H}_0^{\rm m})^2}\bigg] .
\end{equation} 
Using the analyticity of the integrand, the Bromwich contour ${C_\gamma}$ can be deformed into the contour ${\mathcal C_+}$ with the opposite direction, and one finds Eq.~\eqref{eq:Deltab11_v2}.

\section*{CONCLUSIONS}

The main results of this paper are the expression of the scattering phase-shift in Eq.~\eqref{eq:def_delta} as a function of the waveguide frequency (contained implicitly in ${G_0^{\rm rel}}$), the expression of the cumulant in Eq.~\eqref{eq:Deltab11_v4} and the derivation of the Beth-Uhlenbeck formula in Eq.~\eqref{eq:Deltab11_v6}. The discussion of the existence of bound states and the values of the binding energies in atomic waveguides needed to compute Eq.~\eqref{eq:Deltab11_v6} are rather technical and are given in Ref.~\cite{Kri15}. Hopefuly, the simplest way to compute the culmulant ${b_{1,1}}$ (or ${b_2}$ for bosons) is to use directly the expression in Eq.~\eqref{eq:Deltab11_v4} by modifying the contour ${\mathcal C_+}$ in the complex plane with incoming and outgoing directions different from the horizontal axis. In this way, one can avoid the computation of the binding energies.

We have used two methods to derive an explicit formula for the second order virial coefficient for Bose and Fermi atomic gases in atomic waveguides in the vicinity of a Feshbach resonance. We use a quantitative two-channel model which permits us to explicitly take into account the Feshbach mechanism. Our analysis reveals  deep correspondences between two methods which can be extended to the evaluation of higher order in the virial expansion.

As a consequence of the separability of the model, there is formally no change to the expression of the phase-shift or to the usual Beth-Uhlenbeck formula with respect to the homogeneous case and in the absence of Feshbach coupling.  However, as shown in Ref.~\cite{Kri15}, the presence of an atomic waveguide and of inter-channel coupling greatly changes the low-energy properties with respect to the 3D and single channel results. In this way, without using the LDA, our formalism permits us also to obtain quantitative results in regimes where the systems cannot be considered purely 1D or 2D. The existence of a phase-shift in an atomic waveguide is not a general fact. Indeed, in the case where the center of mass and the relative particle motions are not separable (in the presence of anharmonicities, for example) the present analysis does not hold.

Previous studies have focused on the universal character of the virial expansion near resonance. Our results permit us to take into account quantitatively non resonant contributions in two-body scattering. This thus provides a quantitative basis for analysis of experiments where one expects deviations from universal laws due to the presence of an atomic waveguide and/or a finite detuning from resonance.  

One can note that the present formalism can be adapted to evaluate the spectral densities of atomic gases in waveguides including the closed channel by expanding the self-energy in terms of the fugacity \cite{Nga13,Bar14,Sun15}.

\appendix*

\section{Gas without interaction}
\label{sec:non_interacting}
\subsection{Cumulants}

We consider the limit of a large box in the free direction(s) ${(L\to \infty)}$. The mean total number of species ${(\eta \in \{{\rm b},\uparrow,\downarrow,{\rm m}\})}$ is given at equilibrium by
\begin{equation}
\langle \hat{N}_\eta\rangle (\mu_\eta,T)= \sum_{\alpha} n_\eta(\epsilon_\eta^\alpha-\mu_\eta) .
\label{eq:N_eta_free}
\end{equation}
In Eq.~\eqref{eq:N_eta_free}, depending on the statistics considered, ${n_{\eta}}$ is the Bose-Einstein or the Fermi-Dirac distribution (the molecules considered here are bosons). In this section, we formally assume the thermodynamic equilibrium between atoms and molecules without any interaction by imposing the relations in Eq.~\eqref{eq:equilibrium_B} or in Eq.~\eqref{eq:equilibrium_F}. For convenience, in what follows we use the variable ${\chi_\eta=1}$ (${\chi_\eta=-1}$) for the bosonic (fermionic) species ${\eta}$ [see Eq.~\eqref{eq:statistics}]. We can thus write the occupation number in the following form:
\begin{equation}
n_{\eta}(E-\mu_\eta)=\frac{z_\eta e^{-\beta E}}{1-\chi_\eta z_\eta e^{-\beta E}} .
\label{eq:n_eta}
\end{equation}
To achieve the virial expansion for the non-interacting gas we expand Eq.~\eqref{eq:n_eta} in the power of the fugacity:
\begin{equation}
\label{eq:expansion_Bose_Fermi}
n_{\eta}(E-\mu_\eta) =  \sum_{n\ge 1 } \left(\chi_\eta \right)^{n+1} \left(z_\eta e^{-\beta E} \right)^n.
\end{equation}
Using Eq.~\eqref{eq:expansion_Bose_Fermi} in Eqs.~\eqref{eq:N_eta_free} and \eqref{eq:Nb_plus_Nm}, one obtains the cumulants of the non-interacting Bose gas, denoted ${b_{n}^{(0)}}$,
\begin{equation}
b_{2p+1}^{(0)} = \frac{f_D^{\rm b}(2p+1)}{Q} ;  
b_{2p}^{(0)} = \frac{f_D^{\rm b}(2p)+f_D^{\rm m}(p)e^{-p \beta E_{\rm m}}}{Q},  
\label{eq:cumulants_free_B}
\end{equation}
where the function ${f_D^\eta}$ is given by
\begin{equation}
f_D^\eta(n)=\frac{L^D}{\lambda_\eta^D \left[2\sinh\left({n\beta\hbar{\omega}/2}\right) \right]^{3-D}} \times \frac{1}{n^{1+D/2}} .
\label{eq:fDeta}
\end{equation}
In Eq.~\eqref{eq:fDeta}, ${\lambda_\eta}$ is the de Broglie wavelength of particle ${\eta}$:
\begin{equation}
\lambda_\eta = \sqrt{\frac{2\pi \beta \hbar^2}{m_\eta}}.
\end{equation} 
Similarly, the expressions of the cumulants for a non-interacting Fermi gas denoted ${b_{n,p}^{(0)}}$ are deduced from  Eqs.~\eqref{eq:expansion_Bose_Fermi},~\eqref{eq:Nup_plus_Nm} and \eqref{eq:Ndo_plus_Nm}:
\begin{align}
&b_{n,0}^{(0)} = (-1)^{n+1}\frac{f_D^\uparrow(n)}{Q} \quad ; \quad b_{0,n}^{(0)} = (-1)^{n+1}\frac{f_D^\downarrow(n)}{Q}
\label{eq:cumulants_free_F}\\ 
&b_{n,n}^{(0)} = \frac{f_D^{\rm m}(n)}{Q} e^{-n \beta E_{\rm m}} \quad ; \quad b_{n,p}^{(0)}=0 \quad\text{otherwise}.
\end{align}
In the particular case where ${n=p=1}$, one finds a result independent of the trap frequency:
\begin{equation}
b_{1,1}^{(0)} = \left(1+\frac{m_\downarrow}{m_\uparrow}\right)^{D\over2} e^{-\beta E_{\rm m}}
\label{eq:b110_3D_F}
\end{equation}

\subsection{$3D$ dominated regime}

In the limit where ${\beta\hbar{\omega} \ll 1}$, the particles occupy high-energy transverse modes of the waveguide and the system thus has a 3D character. In this 3D-dominated regime, the function in Eq.~\eqref{eq:fDeta} can be approximated by
\begin{equation}
f_D^\eta(n) \simeq \frac{L^D}{\lambda_\eta^D (\beta\hbar{\omega})^{3-D}n^{4-D/2}} .
\end{equation}
The mean number of particles for each species is then
\begin{equation}
\langle \hat{N}_\eta \rangle 
= \frac{ \chi_\eta L^D g_{3-\frac{D}{2}}( \chi_\eta z_\eta)}
{\lambda_\eta^D (\beta \hbar {\omega})^{3-D}} ,
\label{eq:free_species_LDA}
\end{equation}
where ${g_\alpha(t)=\sum_{n\ge 1} t^n/n^\alpha}$. One can verify that, as expected, it is possible to derive Eq.~\eqref{eq:free_species_LDA} after spatial integration of the 3D density approximated in the LDA by using 
\begin{equation}
\langle \hat{N}_\eta \rangle  = \int  d^3r \int \frac{d^3 k}{(2\pi)^3} n_\eta\left(\frac{\hbar^2 k^2}{2 m_\eta}-\mu_\eta -V_{\rm trap}(\mathbf r) \right),
\end{equation}
where ${V_{\rm trap}}$ is the transverse harmonic potential of frequency ${\omega}$. 

In the 3D dominated regime 
$
{\frac{f_D^\eta(n)}{Q} \sim \frac{\lambda^D}{\lambda_\eta^D n^{4-D/2}}}
$
and the second order cumulant for a Bose gas is 
\begin{equation}
b_{2}^{(0)} \simeq \frac{1}{2^{4-D/2}} + 2^{D\over 2} e^{-\beta E_{\rm m}},   
\label{eq:second_order_free_3D_B}
\end{equation}
and for a Fermi gas, the second order cumulants are
\begin{equation}
b_{2,0}^{(0)} \sim -\frac{1}{2^{4-D/2}} \ ; \ b_{0,2}^{(0)} \sim   -\frac{1}{2^{4-D/2}} \left(\frac{m_\downarrow}{m_\uparrow}\right)^{D\over2} .
\label{eq:second_order_free_3D_F}
\end{equation}

These can be compared with the virial coefficients of the homogeneous 3D problem without molecules: for a Bose gas ${b_2^{\rm (hom)}=1/(4\sqrt{2})}$ and for a two-component Fermi gas ${b_{2,0}^{\rm (hom)}=b_{0,2}^{\rm (hom)}=-1/(4\sqrt{2})}$. These results coincide with the analysis in Ref.~\cite{Liu09} adapted for a $D$-dimensional waveguide, which gives ${b_2=2^{(D-3)/2}b_2^{\rm (hom)}}$ and for ${m_\uparrow=m_\downarrow}$, ${b_{2,0}=b_{0,2}=2^{(D-3)/2}b_{2,0}^{\rm (hom)}}$.

We introduce the total number of atoms, including the bounded atoms in the molecular state: ${N=\langle \hat{N}_b \rangle + 2 \langle \hat{N}_{\rm m} \rangle}$ for bosons and ${N=\langle \hat{N}_\uparrow \rangle + \langle \hat{N}_\downarrow \rangle  + 2 \langle \hat{N}_{\rm m} \rangle}$ for fermions. In the 3D-dominated regime, we see from Eq.~\eqref{eq:free_species_LDA} that the dimensionless quantity
\begin{equation}
N \left(\frac{\lambda}{L}\right)^{D} \times \left(\beta \hbar {\omega} \right)^{3-D} 
\end{equation}
 has a finite limit when ${N\to \infty}$, ${L\to \infty}$, and ${{\omega} \to 0}$. This defines the thermodynamic limit of the problem where the LDA can be applied.

\subsection{Low dimensional regime}

In the low-dimensional regime  the temperature is sufficiently small that the particles populate only the lowest transverse state of the harmonic waveguide. This regime is achieved in the limit ${\beta \hbar {\omega} \gg 1}$, and one can use the approximation from \eqref{eq:fDeta},
\begin{equation}
f_D^\eta(n) \sim \frac{L^D}{\lambda_\eta^D}   \times \frac{\exp\left(- n \beta E_0 \right)}{n^{1+D/2}} .
\label{eq:fDeta_low_D}
\end{equation}
In this regime the mean number of particles for each species is given by
\begin{equation}
\langle \hat{N}_\eta \rangle (\mu,T)  = \chi_\eta \frac{L^D}{\lambda_\eta^D} g_{\frac{D}{2}}\left( \chi_\eta e^{\beta \mu_\eta'}\right) ,
\end{equation}
where for ${\eta={\rm b}, \uparrow, \downarrow}$ we have introduced the shifted chemical potential ${\mu_\eta' = \mu_\eta -E_0}$ and for  ${\eta={\rm m}}$, ${\mu_{\rm m}' = \mu_{\rm m} - E_0 - E_{\rm m}}$. We thus recover the same result as for a strictly $D$-dimensional homogeneous system with a chemical potential shifted by the zero point motion of the transverse potential. The cumulant ${b_n^{(0)}}$ for a Bose gas [${b_{n,p}^{(0)}}$ for a Fermi gas] differs from those of the strictly homogeneous $D$ dimensional system by a factor ${\exp(-n \beta E_0)}$ [${\exp(- (n+p)\beta E_0)}$]. The thermodynamic limit in the low-dimensional regime  ${\langle \hat{N}_\eta \rangle \to+\infty}$, ${L\to+\infty}$ is obtained as expected for fixed values of ${\beta}$, and ${\mu'_\eta}$.


\begin{thebibliography}{0}
\expandafter\ifx\csname natexlab\endcsname\relax\def\natexlab#1{#1}\fi
\expandafter\ifx\csname bibnamefont\endcsname\relax
  \def\bibnamefont#1{#1}\fi
\expandafter\ifx\csname bibfnamefont\endcsname\relax
  \def\bibfnamefont#1{#1}\fi
\expandafter\ifx\csname citenamefont\endcsname\relax
  \def\citenamefont#1{#1}\fi
\expandafter\ifx\csname url\endcsname\relax
  \def\url#1{\texttt{#1}}\fi
\expandafter\ifx\csname urlprefix\endcsname\relax\def\urlprefix{URL }\fi
\providecommand{\bibinfo}[2]{#2}
\providecommand{\eprint}[2][]{\url{#2}}

\end{thebibliography}


\begin{thebibliography}{99}

\bibitem{Chi10} C. Chin, R. Grimm, P. Julienne, and E. Tiesinga, 
\href{http://link.aps.org/doi/10.1103/RevModPhys.82.1225}
{Rev. Mod. Phys. {\bf 82}, 1225 (2010)}.

\bibitem{Ho04a} T.-L. Ho, 
\href{http://link.aps.org/doi/10.1103/PhysRevLett.92.090402}
{Phys. Rev. Lett. {\bf 92}, 090402 (2004).}

\bibitem{Nas10a} S. Nascimb\`{e}ne, N. Navon, K. J. Jiang, F. Chevy and C. Salomon, 
\href{http://dx.doi.org/10.1038/nature08814}
{Nature {\bf 463}, 1057 (2010).}

\bibitem{Nas10b} 
S. Nascimb\`{e}ne , N. Navon, F. Chevy and C. Salomon, 
\href{http://iopscience.iop.org/1367-2630/12/10/103026}
{New J. Phys. {\bf 12}, 103026 (2010).}

\bibitem{Hor10} 
M. Horikoshi, S. Nakajima,  M. Ueda, T. Mukaiyama,
\href{http://dx.doi.org/10.1126/science.1183012}
{Science {\bf 327}, 442 (2010).} 

\bibitem{Ku12} M.J.H. Ku, A.T. Sommer, L.W. Cheuk, M.W. Zwierlein, 
\href{http://dx.doi.org/10.1126/science.1214987}
{Science {\bf 335}, 563 (2012).}

\bibitem{Hou12} K. Van Houcke, F. Werner, E. Kozik, N. Prokof’ev,	B. Svistunov, M. J. H. Ku, A. T. Sommer, L. W. Cheuk, A. Schirotzek and M. W. Zwierlein, 
\href{http://dx.doi.org/10.1038/nphys2273} 
{Nature Physics {\bf 8}, 366 (2012).}    

\bibitem{Rem13} B. S. Rem, A. T. Grier, I. Ferrier-Barbut, U. Eismann, T. Langen, N. Navon, L. Khaykovich, F. Werner, D.S. Petrov, F. Chevy, and C. Salomon,
\href{http://link.aps.org/doi/10.1103/PhysRevLett.110.163202}
{Phys. Rev. Lett. {\bf 110}, 163202 (2013).}

\bibitem{Lau14} 
S. Laurent, X. Leyronas, and F. Chevy,
\href{http://link.aps.org/doi/10.1103/PhysRevLett.113.220601}
{Phys. Rev. Lett. {\bf 113}, 220601 (2014).}

\bibitem{Mak14} P. Makotyn,	C. E. Klauss,	D.L. Goldberger,	E. A. Cornell, and D. S. Jin, 
\href{http://dx.doi.org/10.1038/nphys2850}
{Nature Physics {\bf 10}, 116 (2014).}

\bibitem{Fer10} F. Ferlaino and R. Grimm, 
\href{http://dx.doi.org/10.1103/Physics.3.9}
{Physics {\bf 3}, 9 (2010)}.

\bibitem{Hua87}  K. Huang, {\it Statistical Mechanics}, 2nd ed. (John Wiley \& Sons, New York, 1987).

\bibitem{Ho04b} T.-L. Ho and E.J. Mueller,  
\href{http://link.aps.org/doi/10.1103/PhysRevLett.92.160404}
{Phys. Rev. Lett. {\bf 92}, 160404 (2004).}

\bibitem{Liu09} 
X.-J. Liu, H. Hu, and P.D. Drummond, 
\href{http://link.aps.org/doi/10.1103/PhysRevLett.102.160401}
{Phys. Rev. Lett. {\bf 102}, 160401 (2009).}

\bibitem{Liu10a} X.-J. Liu, H. Hu and P.D. Drummond,  
\href{http://dx.doi.org/10.1103/PhysRevA.82.023619}
{Phys. Rev. A {\bf 82}, 023619 (2010).}

\bibitem{Liu10b} X.-J. Liu and H. Hu,  
\href{http://dx.doi.org/10.1103/PhysRevA.82.043626}
{Phys. Rev. A {\bf 82}, 043626 (2010).}

\bibitem{Liu10c} 
X.-J. Liu, H. Hu, and P.D. Drummond, 
\href{http://dx.doi.org/10.1103/PhysRevB.82.054524}
{Phys. Rev. B {\bf 82}, 054524 (2010).}

\bibitem{Ley11} X. Leyronas, 
\href{http://link.aps.org/doi/10.1103/PhysRevA.84.053633}
{Phys. Rev. A {\bf 84}, 053633 (2011).}

\bibitem{Dai12} K. M. Daily and D. Blume, 
\href{http://link.aps.org/doi/10.1103/PhysRevA.85.013609}
{Phys. Rev. A {\bf 85}, 013609 (2012)}.

\bibitem{Rak12} D. Rakshit, K. M. Daily, and D. Blume, 
\href{http://link.aps.org/doi/10.1103/PhysRevA.85.033634}
{Phys. Rev. A {\bf 85}, 033634 (2012).}

\bibitem{Gha12} S. E. Gharashi, K. M. Daily and D. Blume,
\href{http://link.aps.org/doi/10.1103/PhysRevA.86.042702}
{Phys. Rev. A {\bf 86}, 042702 (2012).}

\bibitem{Liu13} X.-J. Liu, 
\href{http://dx.doi.org/10.1016/j.physrep.2012.10.004}
{Phys. Rep. {\bf 524}, 37 (2013)}.

\bibitem{Gao15} 
C. Gao, S. Endo and Y. Castin, 
\href{http://dx.doi.org/10.1209/0295-5075/109/16003}
{Europhys. Lett. {\bf 109}, 16003 (2015).}

\bibitem{Nga15} V. Ngampruetikorn, M.M. Parish, and J.~Levinsen, 
\href{http://link.aps.org/doi/10.1103/PhysRevA.91.013606}
{Phys. Rev. A {\bf 91}, 013606 (2015).}


\bibitem{Cas13} 
Y. Castin, F. Werner, 
\href{http://link.aps.org/doi/10.1139/cjp-2012-0569}
{Can. J. Phys. {\bf 91}, 382 (2013).}

\bibitem{Bar15} 
M. Barth and J. Hofmann, 
\href{http://link.aps.org/doi/10.1103/PhysRevA.92.062716}
{Phys. Rev. A {\bf 92}, 062716 (2015).}

\bibitem{Yan16} Y. Yan and D. Blume, 
\href{http://journals.aps.org/prl/abstract/10.1103/PhysRevLett.116.230401}
{Phys. Rev. Lett. {\bf 116}, 230401 (2016).}

\bibitem{Rei66} A. S. Reiner, 
\href{http://link.aps.org/doi/10.1103/PhysRev.151.170}
{Phys. Rev. {\bf 151}, 170 (1966).}

\bibitem{Bet37} 
E. Beth, G.E. Uhlenbeck,
\href{http://dx.doi.org/10.1016/S0031-8914(37)80189-5}
{Physica {\bf 4}, 915 (1937).}.

\bibitem{QGLD} L. Pricoupenko, H. Perrin, and M. Olshanii (eds.) {\sl Proceedings of the school -- Quantum Gases in Low Dimensions},
\href{http://www.edpsciences.org/articles/jp4/abs/2004/04/contents/contents.html}
{(EDP Sciences, Les Ulis-France, 2004) [J. Phys. IV {\bf 116} (2004)].}

\bibitem{Fen16} K. Fenech, P. Dyke, T. Peppler, M. G. Lingham, S. Hoinka, H. Hu, and C. J. Vale,
\href{http://link.aps.org/doi/10.1103/PhysRevLett.116.045302}
{Phys. Rev. Lett.  {\bf 116}, 045302 (2016).} 

\bibitem{Boe16} I. Boettcher, L. Bayha, D. Kedar, P. A. Murthy, M. Neidig, M. G. Ries, A. N. Wenz, G. Z\"{u}rn, S. Jochim, and T. Enss,
\href{http://link.aps.org/doi/10.1103/PhysRevLett.116.045303}
{Phys. Rev. Lett. {\bf 116}, 045303 (2016).}

\bibitem{Ols98} M. Olshanii, 
\href{http://link.aps.org/doi/10.1103/PhysRevLett.81.938}
{Phys. Rev. Lett. {\bf 81}, 938 (1998)}.

\bibitem{Pet00} D. S. Petrov, M. Holzmann, and G. V. Shlyapnikov, 
\href{http://link.aps.org/doi/10.1103/PhysRevLett.84.2551}
{Phys. Rev. Lett. {\bf 84}, 2551 (2000)}.

\bibitem{Pet01} D. S. Petrov and G. V. Shlyapnikov,
\href{http://link.aps.org/doi/10.1103/PhysRevA.64.012706}
{Phys. Rev. A {\bf 64}, 012706 (2001)}.

\bibitem{Cui12} X. Cui, 
\href{http://link.aps.org/doi/10.1103/PhysRevA.86.012705}
{Phys. Rev. A {\bf 86}, 012705 (2012).}

\bibitem{Nga13} V. Ngampruetikorn, J. Levinsen and M.M. Parish,   
\href{http://link.aps.org/doi/10.1103/PhysRevLett.111.265301}
{Phys. Rev. Lett. {\bf 111}, 265301 (2013).}



\bibitem{Wer09} F. Werner, L. Tarruell, and Y. Castin,
\href{http://www.springerlink.com/index/84j378181w489491.pdf}
{Eur. Phys. J. B {\bf 68}, 401 (2009).}

\bibitem{Jon10} M. Jona-Lasinio, L. Pricoupenko, 
\href{http://link.aps.org/doi/10.1103/PhysRevLett.104.023201}
{Phys. Rev. Lett. {\bf 104}, 023201 (2010).}

\bibitem{Mor11a} C. Mora, Y. Castin, L. Pricoupenko,
\href{http://www.sciencedirect.com/science/article/pii/S1631070510001684}
{C. R. Physique {\bf 12}, 71 (2011).}

\bibitem{Pri11c} L. Pricoupenko, M. Jona-Lasinio, 
\href{http://dx.doi.org/10.1103/PhysRevA.84.062712}
{Phys. Rev. A {\bf 84}, 062712 (2011).}

\bibitem{Tre12} C. Trefzger, and Y. Castin, 
\href{http://link.aps.org/doi/10.1103/PhysRevA.85.053612}
{Phys. Rev. A {\bf 85}, 053612 (2012).}

\bibitem{Kri15} T. Kristensen and L. Pricoupenko, 
\href{http://dx.doi.org/10.1103/PhysRevA.91.042703}
{Phys. Rev. A  {\bf 91}, 042703 (2015).}

\bibitem{Pri13} L. Pricoupenko, 
\href{http://link.aps.org/doi/10.1103/PhysRevLett.110.180402}
{Phys. Rev. Lett. {\bf 110}, 180402 (2013). }

\bibitem{no-pwave} As we neglect the $p$-wave scattering phase-shift, there is no interaction between identical fermions so that ${b_{2,0}=b_{2,0}^{(0)}}$ and ${b_{0,2}=b_{0,2}^{(0)}}$.

\bibitem{Bar14} M. Barth and J. Hofmann, 
\href{http://dx.doi.org/10.1103/PhysRevA.89.013614}
{Phys. Rev. A {\bf 89}, 013614 (2014).}

\bibitem{Sun15} M. Sun and X. Leyronas, 
\href{http://dx.doi.org/10.1103/PhysRevA.92.053611}
{Phys. Rev. A {\bf 92}, 053611 (2015).} 

\end{thebibliography}
\end{document}